\documentclass[twocolumn]{aastex62}
\pdfoutput=1
\nonstopmode
\usepackage{natbib}

\newcommand{\mgii}{\ion{Mg}{2}}
\newcommand{\feii}{\ion{Fe}{2}$^{\ast}$}
\newcommand{\oii}{[\ion{O}{2}]}
\newcommand{\rcsa}{RCSGA032727-132609}

\received{- - -}
\revised{- - -}
\accepted{- - -}

\submitjournal{ApJ}

\shorttitle{Spatially Extended Outflow @ \rcsa}
\shortauthors{Shaban et al.}

\begin{document}

\title{A 30 kpc Spatially Extended Clumpy and Asymmetric Galactic Outflow at $z \sim 1.7$}

\correspondingauthor{Ahmed Shaban, Rongmon Bordoloi}
\email{arshaban@ncsu.edu, rbordol@ncsu.edu}

\author[0000-0002-8858-7875]{Ahmed Shaban}
\affil{Department of Physics, North Carolina State University, Raleigh, NC 27695-8202, USA}

\author[0000-0002-3120-7173]{Rongmon Bordoloi}
\affiliation{Department of Physics, North Carolina State University, Raleigh, NC 27695-8202, USA}

\author[0000-0002-0302-2577]{John Chisholm}
\affiliation{Department of Astronomy, University of Texas at Austin, Austin, TX 78712, USA}

\author[0000-0001-9851-8753]{Soniya Sharma}
\affiliation{Observational Cosmology Lab, NASA Goddard Space Flight Center, Greenbelt, MD 20771, USA}

\author[0000-0002-7559-0864]{Keren Sharon}
\affiliation{Department of Astronomy, University of Michigan, 1085 South University Ave, Ann Arbor, MI 48109, USA}

\author[0000-0002-7627-6551]{Jane R. Rigby}
\affiliation{Observational Cosmology Lab, NASA Goddard Space Flight Center, Greenbelt, MD 20771, USA}

\author[0000-0003-1370-5010]{Michael G. Gladders}
\affiliation{Department of Astronomy and Astrophysics, University of Chicago, 5640 South Ellis Avenue, Chicago, IL 60637, USA }
\affiliation{Kavli Institute for Cosmological Physics at the University of Chicago, USA}

\author[0000-0003-1074-4807]{Matthew B. Bayliss}
\affiliation{Department of Physics, University of Cincinnati, Cincinnati, OH 45221, USA}

\author[0000-0003-0151-0718]{L. Felipe Barrientos}
\affiliation{Instituto de Astrofísica, Pontificia Universidad Catolica de Chile, Santiago, Chile}

\author[0000-0003-0389-0902]{Sebastian Lopez}
\affiliation{Departamento de Astronomía, Universidad de Chile, Casilla 36-D, Santiago, Chile}

\author[0000-0002-1883-4252]{Nicolas Tejos}
\affiliation{Instituto de Física, Pontificia Universidad Católica de Valparaíso, Casilla 4059, Valparaíso, Chile}

\author[0000-0002-7864-3327]{C\'{e}dric Ledoux}
\affiliation{European Southern Observatory, Alonso de Córdova 3107, Vitacura, Casilla 19001, Santiago, Chile}

\author[0000-0001-5097-6755]{Michael K. Florian}
\affiliation{Observational Cosmology Lab, NASA Goddard Space Flight Center, Greenbelt, MD 20771, USA}

\begin{abstract}
We image the spatial extent of a cool galactic outflow with fine structure {\feii} emission and resonant {\mgii} emission in a gravitationally lensed star-forming galaxy at $z = 1.70347$. The {\feii} and {\mgii}  (continuum-subtracted) emissions span out to radial distances of $\sim$14.33 kpc and 26.5 kpc, respectively, with maximum spatial extents of $\sim$21 kpc for {\feii} emission and $\sim$30 kpc for {\mgii} emission. {\mgii} residual emission is patchy and covers a total area of $\sim$184 kpc$^2$, constraining the minimum area covered by the outflowing gas to be $\sim$13\% of the total area. MgII emission is asymmetric and shows $\sim$21\%  more extended emission along the declination direction. We constrain the covering fractions of the {\feii} and {\mgii} emission as a function of radial distance and characterize them with a power law model. The {\mgii} 2803 emission line show two kinematically distinct emission components, and may correspond to two distinct shells of outflowing gas with a velocity separation of $\Delta v \sim$ 400 km/s. By using multiple images with different magnifications of the galaxy in the image plane, we trace the {\feii}, {\mgii} emissions around three individual star-forming regions. In all cases, both the {\feii} and {\mgii} emissions are more spatially extended compared to the star forming regions traced by the {\oii} emission. These findings provide robust constraints on the spatial extent of the outflowing gas, and combined with outflow velocity and column density measurements will give stringent constraints on mass outflow rates of the galaxy.
\end{abstract}


\section{Introduction} \label{sec:intro}
Galactic outflows play an important role in galaxy evolution \citep{somerville2015physical} as they transport baryons from the inter-stellar medium (ISM) of galaxies into their circumgalactic medium \citep[CGM;][]{bordoloi2011radial, tumlinson2017circumgalactic, van2017effect, angles2017cosmic}. This process depletes the gas supply needed to form the next generation of stars in star-forming galaxies and, in extreme cases, can completely quench star-formation in them \citep{man2018star,geach2018violent, hopkins2012stellar, hopkins2014galaxies}. By carrying metals out from the ISM, these outflows can also enrich the intergalactic medium \citep[IGM; ][]{rahmati2016cosmic, ford2016baryon, rupke2018review}. The energy sources driving these outflows can be either star formation (SF) or active galactic nuclei (AGNs) in the galaxy \citep{veilleux2005galactic}. In this work, we will only focus on star-formation driven outflows.

Theorists debate whether star-formation driven galactic outflows are powered by energy from supernovae explosions \citep{chevalier1985wind} or momentum from high-energy photons, and stellar winds, or cosmic rays \citep{murray2005maximum}. The outflows also seem to regulate the star formation and set the mass-metallicity relation \citep{tremonti2004origin}. 
These outflows are also ubiquitous in star-forming galaxies and are complex and multi-phased, by which we mean both ionized and neutral gas with significant dust \citep{veilleux2005galactic, weiner2009ubiquitous, rubin2010galaxies,Martin2013,bordoloi2014dependence, chisholm2015scaling, heckman2015systematic, fiore2017agn,Bordoloi2017FB, cicone2018largely, rupke2018review, schneider2018production}. These different phases of outflow can be detected at many wavelengths, ranging from the X-rays to millimeter and sub-millimeter \citep{rupke2018review}.

While models and simulations require outflows to regulate the star formation within galaxies, constraining the impact of outflows requires estimating the total mass that outflows carry out of galaxies. The rate of mass loss is typically characterized by the mass outflow rate ($\rm{\dot{M}_{out}}$), as,
\begin{equation}\label{eq:mass_out}
    \dot{M}_{out}=\Omega\ C_{f}\ \mu m_p\ N_H\ r\ v_{out},
\end{equation}
where $\Omega$ is the opening angle of the outflowing gas, $C_f$ is the covering fraction or the ratio of the stellar continuum that is covered by the outflow in the context of the ``down-the-barrel" observations, $\mu m_p$ is the mean molecular weight of Hydrogen, $N_H$ is the column density of the outflowing gas, $v_{out}$ is the velocity of the outflowing gas, and $r$ is the distance or the spatial extent of the outflow from the galaxy. The parameters of equation \ref{eq:mass_out} can be observationally constrained in a robust manner from down-the-barrel spectroscopic studies of galactic outflows \citep{chisholm2016robust}. However, the spatial extent ($r$) of the outflow remains largely unconstrained in such works. Therefore, different strategies have been implemented to infer $r$ \citep{rubin2014evidence, heckman2015systematic, bordoloi2016spatially, chisholm2016shining,chisholm2018feeding}. One way to make progress is to use spatially extended emission lines (H$\alpha$, \ion{O}{2}, \mgii, \feii, etc.) that trace the densest phase of the gas in such outflows, to measure the corresponding spatial extent of the outflowing gas \citep{ shapley2003rest, rubin2011low, zhang2016hydrogen,rupke2019100, burchett2020circumgalactic,zabl2021muse}.

This has been done for galaxies in the local universe and galaxies at moderate redshift ($z\approx 0.5$). \cite{rubin2011low} used Keck/LRIS to measure the spatial extent of the wind in the galaxy TKRS4389 ($z\sim 0.47$). The measured extent of the \mgii\ emission doublet 2796, 2803 {\AA} from the wind is $\sim 7$ kpc in one dimension along the slit. The limited slit size will lead to the loss of the signal from the \mgii\ emission from the regions of the galaxy which are not covered by the slit. One can increase the spatial coverage by performing integral field unit (IFU) spectroscopy, which provides a spectrum for each spaxel in the field of view. 

Indeed, \cite{burchett2020circumgalactic} targeted the same galaxy with KCWI/IFU observations and measured a $\sim 31$ kpc spatial extent of the \mgii\ emission. \cite{rupke2019100} studied another galaxy at similar redshift $z\sim 0.46$ using KCWI. They measured the spatial extent of the wind traced by the \oii\ doublet 3726, 3729 {\AA} and detected the emission up to $\sim 100$ kpc, which is the largest measured extent of a galactic outflow. Other \mgii\ IFU observations of extreme galaxies have shown that some galaxies do not have extended \mgii\ outflows, rather strong \mgii\ emission can arise in H II regions within galaxies \citep{chisholm2020optically}.

One of the complexities in tracing outflows using emission lines is the low surface brightness emission in individual galaxies. This makes it hard to detect the emission and localize it to the individual star-forming clumps which might be driving the outflowing gas. To overcome these issues, we can leverage the phenomenon of gravitational lensing and zoom-in on individual star-forming regions in a galaxy \citep{bordoloi2016spatially}. Gravitational lensing stretches sub kpc-scale regions within a galaxy to few arc-seconds on the sky, while conserving the surface brightness of each region. This is very suitable for studying individual star-forming regions within a galaxy, especially when the lensed galaxy have multiple images in the image-plane. One of the conditions for this method to work, is that the lens should have a robust mass model \citep{sharon2012source, sharon2020strong}. Augmented with deep IFU observations, we can obtain very high signal-to-noise-ratio (SNR) observations with large spatial coverage and constrain the properties of galactic outflows at different sizes and scales in the source plane of the galaxies driving them. By using the lens model with the IFU observations, we can trace the 2D maps of the emission lines tracing the outflows to the source plane of the galaxy and obtain a measure for the outflow extent in physical distance. 

In this work, we use VLT/MUSE observations to study the \mgii\ resonant back scattered emission, and the \feii\ fine structure emission in the strong gravitationally lensed galaxy RCSGA 032727-132609 at $z \sim$ 1.703 \citep{wuyts2010bright, wuyts2014magnified}. The \mgii\ and \feii\ emission trace the cool phase of the outflows. We also study the nebular \oii\ emission. We measure the spatial extent of the outflow using the \mgii\ emission. The detailed study of outflow gas kinematics and mass outflow rates will be presented in a separate forthcoming publication (Shaban et al. in prep). 

This paper is organized as follows: \S \ref{sec:obs} describes the MUSE observation; \S \ref{sec:method} describes the method of emission map construction in the image plane, the construction of these maps in the source plane, correction for seeing and lensing shear, and the method of constructing the surface brightness radial profiles; \S \ref{sec:results} describes the results of the analysis. In \S \ref{sec:discuss}, we discuss these results, compare them with the literature, and state the final conclusions of our study. For the rest of this work, we do our calculations assuming a $\Lambda$-Cold Dark Matter ($\Lambda$CDM) cosmology with $H_0 = 70\ {\rm km \ s^{-1} Mpc^{-1}}$, $\Omega_m = 0.3$, and $\Omega_{\Lambda}=0.7$. 

\section{Observations} \label{sec:obs}
\begin{figure*}
    \centering
    \includegraphics[width=\textwidth]{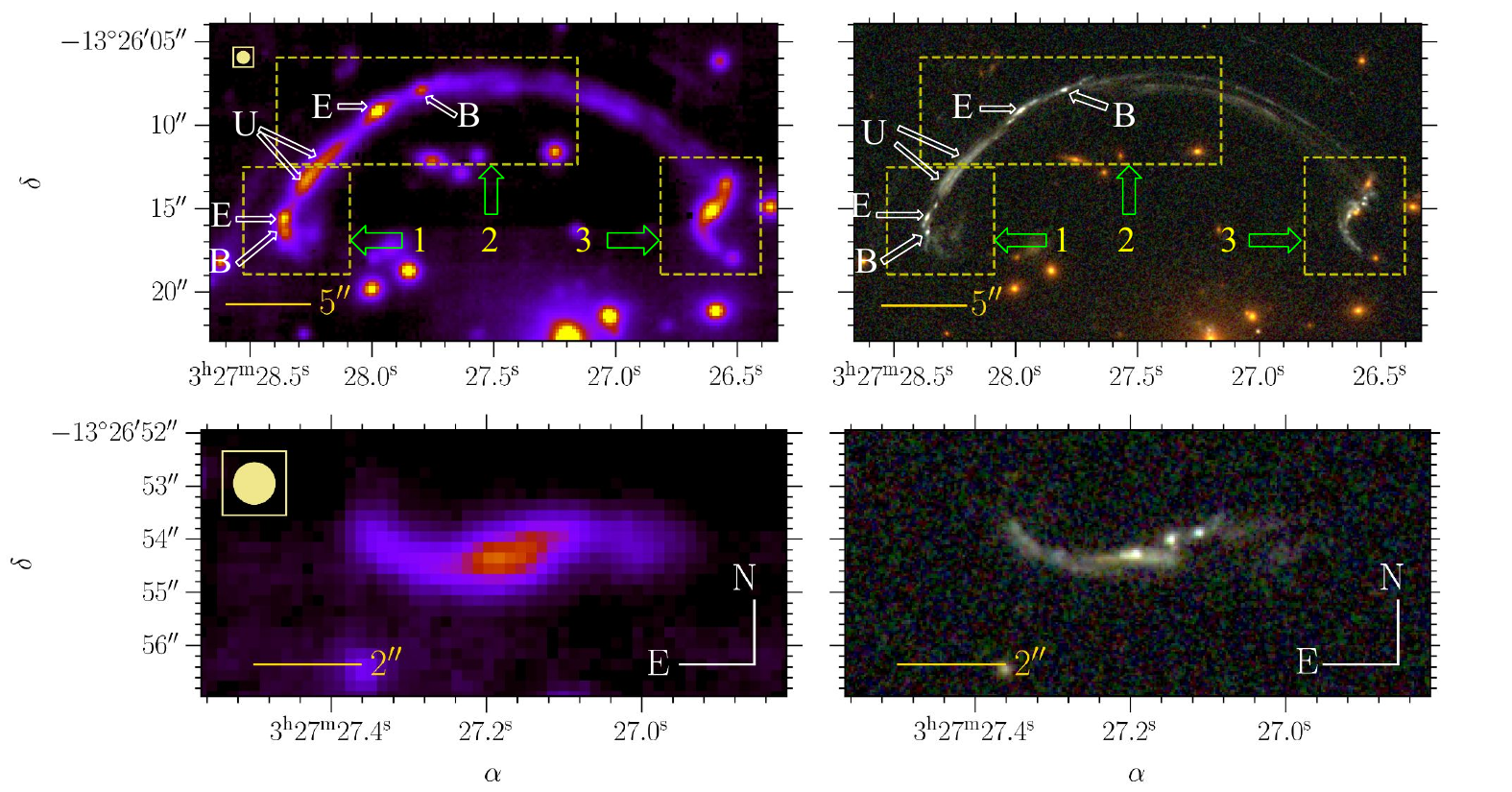}
    \caption{ \rcsa\ is a strong gravitationally lensed galaxy at z $\approx$ 1.703. \textit{Left Columns:} MUSE white light images. \textit{Right Columns:} HST F390W, F606W, and F814W composite images. \textit{Top Rows:} Main arc of the galaxy. Three multiple images are marked with dashed yellow rectangles. The white letters and arrows show the multiply imaged distinct star-forming regions in the image plane. \textit{Bottom Rows:} The counter arc of the galaxy, which shows a less magnified and distorted image of the galaxy. The yellow circles in the left panels represent the maximum seeing during the MUSE observation.}
   \label{fig:muse_hst}
\end{figure*}

The galaxy RCSGA 032727-132609 is a low-metallicity star-forming galaxy at $z= 1.70347\pm 0.00002$ (Section: \ref{sec:method}) and is lensed by the galaxy cluster RCS2 032727-132623 at $z = 0.564$ \citep{wuyts2010bright, wuyts2014magnified,gonzalez2017alma,rigby2018magellanI}. It was discovered in the Second Red Sequence Cluster Survey \citep{gilbank2011red}. The apparent shape of the galaxy consists of the main arc north of the cluster subtending $38''$ and a smaller counter arc south of the cluster subtending $7''$ on the sky \citep[see \autoref{fig:muse_hst}; ] []{wuyts2014magnified}. The main arc consists of three images of the galaxy (images 1, 2, and 3 are denoted by yellow rectangles and yellow numbers in \autoref{fig:muse_hst}) and the counter arc is a $\mathrm{4^{th}}$ image. Image 1 and Image 2 are highly magnified because they are situated near the critical lines in the image plane. Images 1 and 2 represent sub-regions of the galaxy in the image plane. These critical lines correspond to the regions, where there is theoretically infinite magnification. From the best fit model for the lens from \cite{sharon2012source}, the average magnification across the main arc is $25.1_{-2.5}^{+3.2}$, and the average magnification values of the individual images of the main arc are $10.4_{-0.8}^{+1.1}$, $20.6_{-2.2}^{+2.6}$, and $9.7_{-0.9}^{+1.1}$ for images 1, 2, and 3, respectively. The magnification of the counter arc is $3.0_{-0.1}^{+0.2}$. 

This paper focuses on IFU observations of RCSGA 032727-13260 using the VLT/MUSE instrument with program ID: 098.A-0459(A) \citep{lopez2018clumpy}. The observations were taken using the MUSE wide field mode with a spatial sampling of 0.2$''$ per pixel, a field of view of $1^{\prime} \times 1^{\prime}$, and spectral sampling of {1.25 \AA} per pixel and a spectral resolution ($R = \frac{\lambda}{\Delta \lambda}$) of 1770 at 480 nm to 3590 at 930 nm \citep{bacon2010muse}. The total exposure time of the observations is 3.1 hours. During the time of the observations, the maximum atmospheric seeing was 0.8$''$ and the maximum airmass was 1.8. The sky subtraction was applied on the cube using the Zurich Atmospheric Purge (ZAP) algorithm \citep{soto2016zap}. We refer the reader to \cite{lopez2018clumpy} for a detailed description of the observations.
We use \textit{Hubble Space Telescope} WFC3/F390W, WFC3/F606W, and WFC3/F814W imaging of this field (PI: J. Rigby, Proposal ID: 12267) to construct a multi-wavelength composite image of the main arc and the counter arc (Figure \ref{fig:muse_hst}). The observed pivot wavelengths for these filters correspond to galaxy rest-frame wavelengths of 1450 {\AA}, 2176 {\AA}, and 2976 {\AA}, respectively. We use these HST observations to accurately define the astrometry of the MUSE data-cube. We identify three common bright stars in both the MUSE data-cube and the HST images. Then, we match their central pixel coordinates to fix the astrometric offset in the MUSE data-cube. These offsets correspond to a difference in right ascension $\Delta \alpha \approx$ 0.693$''$, and a difference in declination $\Delta \delta \approx$ 3.157$''$, respectively. 

Figure \ref{fig:muse_hst} shows the main arc in the top row and the counter arc in the bottom row, with the multiple images of the galaxy shown in yellow dashed rectangles for both the MUSE white light image (left panels) and the HST composite image (right panels). We follow the naming convention of the star-forming (SF) regions from \cite{bordoloi2016spatially}. There are four SF regions named E, U, B, and G that are multiply imaged in the main arc. They are highlighted with purple arrows in Figure \ref{fig:muse_hst}. Image 2 is the most magnified image, image 1 is the second most magnified image, and both of them represent small individual star-forming regions in the source plane galaxy. Image 3 and the counter arc represent images of the whole galaxy in source plane. The counter arc is the least magnified and least distorted image of the galaxy, and we use it as a representative of our measurements for the whole galaxy.

\section{Methods} \label{sec:method}
We aim to estimate the spatial extent of galactic outflows in this galaxy using the \mgii\ and \feii\ emission lines and compare them to the nebular {\oii} emission. For this purpose, we produce narrowband maps around the emission lines of interest. We use the lens model to reconstruct these emission maps in the source plane of the galaxy and measure the true physical extent of the outflows. We develop and use a python package to do most of this analysis named \texttt{musetools}\footnote{https://github.com/rongmon/musetools}. These steps are described as below.

\subsection{1D Spectral Extraction}
To identify the emission lines of interest, we extract a light weighted 1D spectrum of the main arc of the galaxy. We first select the voxels, which are the data points in the data cube, that cover the main arc. These voxels are summed over the 4600–9350{\AA} wavelength range to create a white light image of the arc. Each voxel is weighted by this white light image and summed in the spatial direction to create a light weighted 1D spectrum of the arc and the counter arc. This method produces a high SNR 1D spectrum of the galaxy and is shown in Figures \ref{fig:muse_FeII_lines} and \ref{fig:stack_spec_mgII_emission}. The emission lines of interest for this study are the \mgii\ emission doublet $\lambda\lambda$ 2796, 2803 {\AA}, five \feii\ fine structure emission lines, and the \oii\ nebular emission doublet $\lambda\lambda$ 2470, 2471 {\AA} \citep{morton2003atomic, leitherer2011ultraviolet}. These lines are summarized in Table \ref{tab:lines_tot}. The \mgii\ and \feii\ emission trace the outflows and the \oii\ emission traces the star-forming regions in the galaxy. The \mgii\ emission doublet shows a P-Cygni profile with the \mgii\ absorption lines. A selection of the specific wavelength intervals for \feii and \mgii\ emission is shown in Figures \ref{fig:muse_FeII_lines} and \ref{fig:stack_spec_mgII_emission}, respectively. The \oii\ emission lines show up as a blended doublet highlighted in green in Figure \ref{fig:muse_FeII_lines}. For the main arc, the average SNR per pixel around the \feii, \oii, and \mgii\ lines are 94, 106, and 105, respectively. For the counter arc, the average SNR per pixel for the same lines are 26, 29, and 28, respectively.

We measure the systemic redshift ($z = 1.70347 \pm 0.00002$) of the galaxy by fitting a double Gaussian to the interstellar medium (ISM) \oii\ emission doublet $\lambda \lambda \ 2470.79, 2471.09$ {\AA} \citep{leitherer2011ultraviolet}.

\subsection{Generating Emission Maps}\label{2D_maps}
\begin{figure*}
    \centering
    \includegraphics[width=\textwidth]{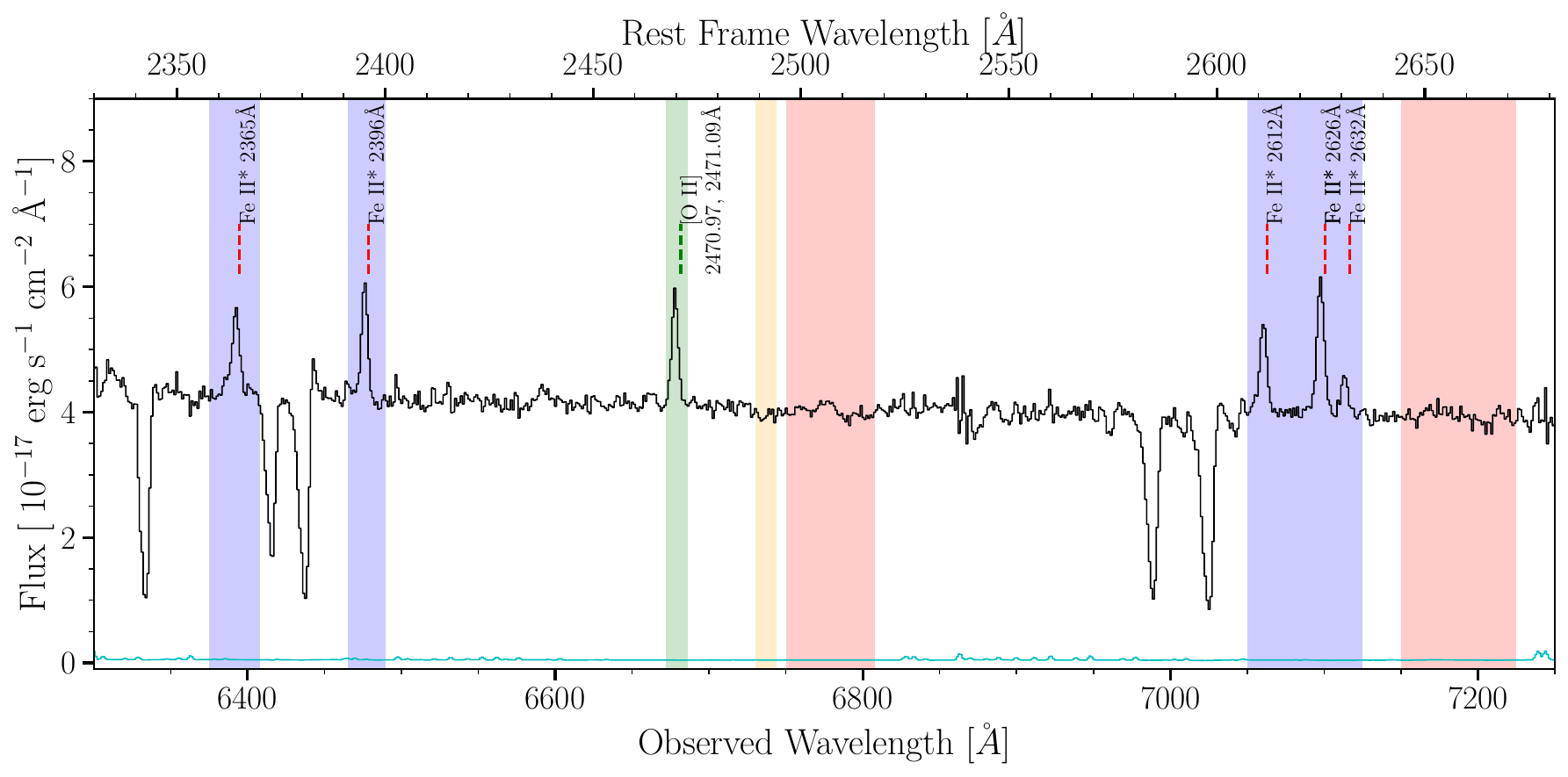}
    \caption{Mean 1D spectrum of the \rcsa\ main arc centered on the \feii\ and \oii\ lines. The \textit{solid black} line represents the flux and the \textit{solid cyan} line represents the corresponding uncertainty on the flux. The shaded boxes show the  wavelength windows used to create the \feii\ emission (\textit{faint blue}), \oii\ emission (\textit{faint green}), and local stellar continuum (\textit{faint red} and \textit{faint orange}) maps, respectively. The width of the stellar continuum windows is equal to the width of the \feii\ and \oii\ emission wavelength windows, respectively.}
    \label{fig:muse_FeII_lines}
\end{figure*}

\begin{figure*}
    \centering
    \includegraphics[width=\linewidth,height=\textheight,keepaspectratio]{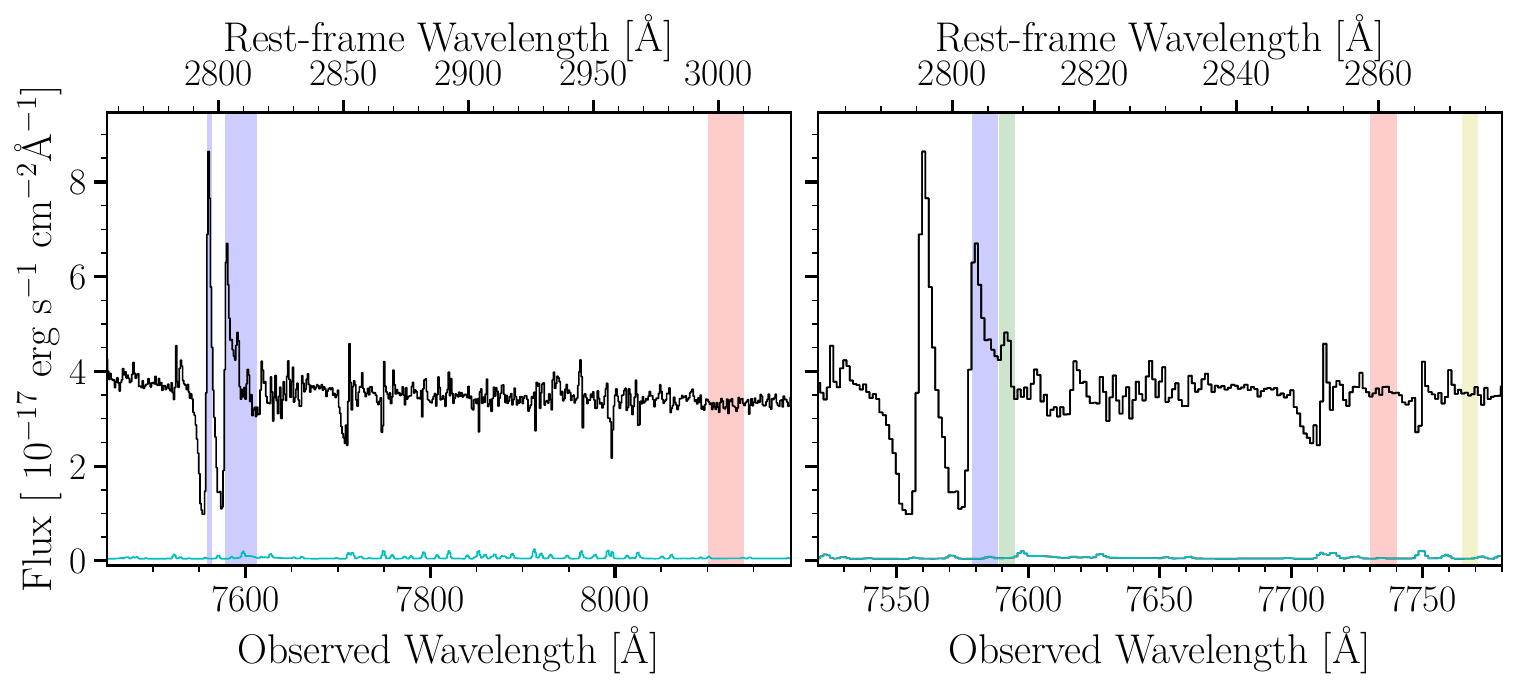}
    \caption{Mean 1D spectrum of the main arc containing the \mgii\ doublet. The \textit{solid black} line shows the flux, and the \textit{solid cyan} line shows the error on the flux. \textit{Left:} The \textit{faint blue} regions represent the selected wavelength window around the \mgii\ emission lines. The \textit{faint red} regions represent the selected wavelength window for the corresponding continuum region. \textit{Right:} The selection of the wavelength windows for the primary and secondary emission peaks of the \mgii\ 2803 emission line. The \textit{faint blue} and \textit{faint red} regions represent the emission and continuum windows for the primary peak, respectively. The \textit{faint green} and \textit{faint yellow} regions represent the emission and continuum windows for the secondary peak, respectively.}
    \label{fig:stack_spec_mgII_emission}
\end{figure*}

To produce narrowband maps around an emission line of interest, we follow the following procedure.
\begin{itemize}
    \item We select a wavelength window $\Delta \lambda$ over which a narrowband image is to be created (see Table \ref{tab:sigma_values}, Figure \ref{fig:muse_FeII_lines}, and Figure \ref{fig:stack_spec_mgII_emission}).
    
    \item We sum all the flux voxels in that wavelength window $\Delta \lambda$ and multiply by the wavelength width per pixel $\delta \lambda$ (${\rm \delta \lambda = 1.25 \AA}$ for MUSE) and divide by the angular area of each pixel (pixel area $\Delta xy$ = $(2'')^2$ ) to create a narrowband surface brightness image of the emission line and the underlying continuum of the galaxy. 
    \begin{equation}
      SB_{(i,j)} = \frac{\delta \lambda}{\Delta xy} \times \displaystyle\sum_{l=\lambda_{min}}^{l=\lambda_{max}} f_{(l,i,j)} 
    \end{equation}
    where $SB_{(i,j)}$ is the surface brightness at the $(i,j)$th pixel measured in units of erg s$^{-1}$ cm$^{-2}$ arcsec$^{-2}$, $f_{(l,i,j)}$ is the flux density at the $(l,i,j)$ voxel measured in units of erg s$^{-1}$ cm$^{-2}$ \AA$^{-1}$.
    
    \item We create a pure continuum emission map by specifying another wavelength interval redward of the emission lines (Table \ref{tab:sigma_values}) that has a wavelength window of identical width ($\Delta \lambda$) as the one chosen for the previous step. These voxels are summed to create a pure continuum surface brightness map of the galaxy. As young stellar populations have very featureless continuum regions in the rest frame 2000-3000 {\AA} range, this method creates robust continuum maps \citep{leitherer1999starburst99}.
    
    \item We subtract the emission+continuum images with the pure continuum images to produce a (continuum-subtracted) emission maps.
\end{itemize}

\begin{deluxetable}{ccc}
\tablecolumns{3}
\tablewidth{0pt}
\tablecaption{Absorption and emission lines used in this work.}
\tablehead{
\colhead{Transition\tablenotemark{a}}&
\colhead{$\lambda$\tablenotemark{b}}&
\colhead{Type}}
\startdata
\mgii\ & 2796.351 & Resonant abs/ems \\ 
 & 2803.528 & Resonant abs/ems \\
\feii\ & 2365.552 & Fine-structure ems \\
 & 2396.355 & Fine-structure ems \\
 & 2612.654 & Fine-structure ems \\
 & 2626.451 & Fine-structure ems \\
 & 2632.108 & Fine-structure ems \\
\oii\ & 2470.97 & Nebular Emission \\
 & 2471.09 & Nebular Emission \\
\enddata
\vspace{-0.2cm}
\label{tab:lines_tot}
\tablenotetext{a}{Atomic data from \cite{morton2003atomic} and \cite{leitherer2011ultraviolet}}
\tablenotetext{b}{Vacuum wavelength in \AA.}
\end{deluxetable}

\begin{deluxetable}{ccccc}
\tabletypesize{\footnotesize}
\tablewidth{\columnwidth}
\tablecaption{ $1\sigma$ Surface brightness background level for each map in units of $10^{-19}$ erg s$^{-1}$ cm$^{-2}$ arcsec$^{-2}$.}
\tablehead{
\colhead{Transition} &
\colhead{E+C\tablenotemark{a}} &
\colhead{C\tablenotemark{b}} &
\colhead{E}\tablenotemark{c} &
\colhead{$\Delta \lambda$ [\AA]\tablenotemark{d}}
}
\startdata
{\oii} & 2.91 & 3.66 & 5.02 & 14 \\ 
{\feii} & 9.82 & 9.96 & 11.76 & 133 \\ 
Full {\mgii} & 11.09 & 6.73 & 11.4 & 39 \\ 
{\mgii} 2803 Primary & 3.7 & 7.6 & 9.29 & 10 \\ 
{\mgii} 2803 Secondary & 4.13 & 4.59 & 7.41 & 6 \\ 
\enddata
\vspace{-0.2cm}
\label{tab:sigma_values}
\tablenotetext{a}{Emission+Continuum.}
\tablenotetext{b}{Continuum.}
\tablenotetext{c}{Continuum-Subtracted Emission.}
\tablenotetext{d}{Width of each wavelength window in {\AA}.}
\end{deluxetable}

We define an average background surface brightness noise level for each map to quantify the statistical significance of individual emission features. We select a $8'' \times 8''$ square region north-east of the main arc where there are no galaxies or foreground stars. In each narrowband map, we compute the standard deviation in surface brightness within this square and use it as the average background surface brightness noise level. The $1\sigma$ background surface brightness levels for the different narrowband maps are summarized in Table \ref{tab:sigma_values}. We use these background surface brightness levels to quantify the statistically significant emission in the rest of the paper.

\subsection{Source Plane Reconstruction}\label{sec:reconstruction}
All the 2D images from the data cube are in the image plane. To reconstruct the source plane emission maps, we use the software package LENSTOOL \footnote{http://projects.lam.fr/projects/lenstool} \citep{kneib1996hubble,jullo2007bayesian,jullo2009multiscale} using the lensing model from \cite{sharon2012source, lopez2018clumpy}. Specifically, we used the direct reconstruction (cleanlens task) in LENSTOOL to convert the image plane fluxes to the source plane with ray-tracing provided by our best-fit lensing mass model. We preserve the surface brightness to accurately reconstruct the surface brightness distribution of different images in the source plane. For the highly magnified images 1 and 2, we use the cleanlens task of LENSTOOL with a grid over-sampling parameters of \texttt{ech} = 10 on the image and \texttt{sech} = 8 on the source plane to obtain a source pixel size of $0.025''$ (0.21 kpc). For image 3 and the counter arc, we choose a sub-sampling parameters of \texttt{ech} = 5 on the image and \texttt{sech} = 3 on the source plane to create the reconstructions. This resulted in a pixel size of $0.067''$ (0.56 kpc) in the source plane. We also propagate the uncertainties on the lens model to all our measured distances and areas in the source plane.

The maximum atmospheric seeing at the time of the observation was $0.8''$. This corresponds to 4 spatial pixels in the image plane based on MUSE spatial resolution. We express it analytically as a 2D normalized Gaussian with a FWHM = $0.8''$ or 4 pixels in the image plane. To account for the effects of the seeing in the source plane, we inject a 2D Gaussian at the central regions of images 1, 2, 3, and the counter arc in the image plane. We then reconstruct that 2D Gaussian in the source plane to account for atmospheric seeing in our observations. These reconstructions are shown as filled yellow ellipses in all source-plane images.

\subsection{Lens Shear and Seeing Correction}\label{sec:shearing_lensing_correction}
As lensing shear is different along each axis, it shears the seeing differently in the source plane. So instead of a symmetric seeing disk in the image plane, one needs to account for an elliptical smear in the source-plane. Both the effect of seeing and the effect of this shear needs to be accounted for to ascertain the true spatial asymmetry of the {\mgii} emission. We perform a suite of simulations to account for this effect on the measured spatial extent in both $\alpha$ and $\delta$ directions. We inject a series of 2D Gaussians, each with FWHMs ranging from 4 pixels to 50 pixels (pixel physical size $= 0.56$ kpc), in the source plane at the location of the counter arc. We then convolve them with the reconstructed seeing in the source plane at the location of the counter arc. We measure the observed x-extent and y-extent after the convolution. This allows us to constrain the impact of seeing in the source-plane of the counter arc, as any variation from the circular shape of the injected 2D Gaussian is due to seeing and lensing shear. This effect is very evident for a 2D Gaussian with small FWHM ($\sim$ 4 pixels) and is almost negligible at FWHM of 50 pixels. Using these simulations, we can correct our measurements for any asymmetry owing to atmospheric seeing and lensing shear. We invert the relationship between the observed and true x and y extents of the simulated images, and compute the true size of the structure for any observed x or y extents.

\subsection{Radial Profiles}\label{sec:Method_Radial_Profiles}
To quantify the spatial extent of the nebular emission and the cool galactic outflow, we need to constrain the {\oii}, {\feii}, and {\mgii} emission surface brightness radial profiles.

We first reconstruct all the image-plane emission maps in the source plane of the galaxy. For each image or star-forming region, we select the pixel with the maximum surface brightness as the center. We compute the mean surface brightness profile in the source plane as a function of physical distance with the center defined above. We use jackknife re-sampling from \texttt{astropy} \citep{astropy2013astropy,astropy2018astropy} to quantify the uncertainty in mean surface brightness of each radial bin. In short-- in each radial bin, a pixel is randomly excluded and randomly replaced with one of the remaining pixels. We compute the mean surface brightness and repeat the step until each pixel has been excluded once at least. The 16th and 84th percentile of the final mean surface brightness distribution gives us the 1$\sigma$ uncertainty of mean surface brightness in each radial bin.

We parametrize the emission surface brightness radial profiles with an exponential for the inner region of the profile plus a power law for the outer region as follows: 
\begin{equation}\label{eq:radial_profile_model}
    \rm{ SB(r) = 
        I_{0,1} e^{-(\frac{r}{r_0})} +
        I_{0,2} \left(\frac{r}{r_b}\right)^{\beta} }
\end{equation}
where $I_{0,1}$ is the surface brightness intensity at $r = 0$, $r_{0}$ is a scale radius for the exponential, $I_{0,2}$ is the surface brightness intensity at $r = r_b$, $r_b$ is the characteristic radius for the power law, and $\beta$ is the index of the power law. Then, we convolve equation \ref{eq:radial_profile_model} with the corresponding reconstructed seeing for each region in the source plane. We fit this convolved model using the Markov chain Monte Carlo (MCMC) sampling using the python package $\texttt{EMCEE}$ \footnote{https://emcee.readthedocs.io/} \citep{foreman2013emcee}. The best fitting parameters using the model described here are summarized in Table \ref{table:radial_parameters}.1.

\subsection{Covering Fraction}\label{sec:covering_fraction}
Using the source-plane reconstructed emission maps of the counter arc, we constrain the observed spatial incidence of the {\oii}, {\feii} and {\mgii} emitting regions around the galaxy. We quantify this as the emission covering fraction $C_f(r)$:
\begin{equation}
    C_f(r) = \frac{N_{(>3\sigma)}(r)}{N_{\rm total}(r)},
\end{equation}
where $N_{(>3\sigma)}$(r) is the number of pixels within a radial bin, that are detected at higher than $3\sigma$ significance relative to the background, and $N_{\rm total}$(r) is the total number of pixels in the same bin. We use the Wilson score interval\footnote{https://github.com/rongmon/rbcodes/} to constrain the confidence intervals of $C_f(r)$. $C_f(r)$ effectively quantifies the fraction of the total area around the galaxy within a radial bin, where \mgii\ emission is detected. 

\section{Results} \label{sec:results}

In the following sections we present the spatial extent of \feii, \oii, and \mgii\ emission detected around \rcsa\ and quantify the spatial radial profiles and covering fractions as a function of galactocentric radius. We further measure the maximum spatial extent of emission as the maximum separation between significant emission spaxels along the x-axis or y-axis. 

\subsection{\oii\ Nebular Emission}
We first study the spatial extent of the nebular \oii\ emission traced by the emission doublet at $\lambda \lambda$ 2470.79, 2471.09{\AA}. We create an emission map around the doublet as described in Section \ref{sec:method}. 
\subsubsection{\oii\ emission in the image plane}
Figure \ref{fig:muse_OII_FeII_MgII_map_full}, left panels show the \oii\ continuum-subtracted emission maps in the image plane of the main arc (top panel) and the counter arc (bottom panel), respectively. The white contours show the 3$\sigma$ surface brightness significance level of the \oii\ (continuum-subtracted) emission. \oii\ emission features are comparable to stellar continuum light in spatial extent. This suggests that in this galaxy, \oii\ emission is not spatially extended beyond the stellar continuum. This is in contrast to what is seen in \cite{rupke2019100}, where the \oii\ emission doublet $\lambda \lambda$ 3726, 3729 {\AA} could be seen extending out to 100 kpc from a low-$z$ ($z\sim 0.5$) star-bursting galaxy.

\subsubsection{\oii\ emission in the source plane}\label{sec:result_OII_source}
We reconstruct the (continuum-subtracted) \oii\ emission in the source plane to quantify the spatial extent. Figure \ref{fig:source_OII_FeII_MgII_full_CARC}, left panel, shows the source plane reconstruction of the \oii\ surface brightness for the counter arc up to 3$\sigma$ significance level. As the counter arc represents the entire galaxy in the image plane, the source plane reconstructed image covers the full spatial extent of the galaxy. We also plot the 3$\sigma$ surface brightness contours over the source-plane reconstruction of the HST image (bottom panel). The \oii\ emission contours follow the stellar light in the HST emission very closely. This suggests that the \oii\ nebular emission regions are the same regions emitting light in the HST reconstruction.

We measure the maximum spatial extent of \oii\ emission along the x- and y-axes, and the surface brightness radial profile in the counter arc. Figure \ref{fig:spatial_XY_extent_src}, left column, shows these x- and y-extents. We see that these distances extend beyond the seeing in the source plane. The measured observed values are summarized in Table \ref{tab:CARC_MgII_distances}. These distances correspond to $\Delta \alpha\ \mathrm{or}\ \Delta x = 13.5_{-0.2}^{+0.3}$ kpc and $\Delta \delta\ \mathrm{or}\ \Delta y = 9.2_{-0.3}^{+0.4}$ kpc after lens shear+seeing correction.

We also characterize the azimuthally averaged surface brightness as a mean 1D radial emission profile in the left panel of Figure \ref{fig:source_radial_OII_FeII_MgII} as described in Section \ref{sec:Method_Radial_Profiles}. We define the center of the radial profile as the brightest pixel of the emission map at the counter arc in the source plane. In each radial bin, we compute the mean surface brightness of all pixels above $3\sigma$ significance level, shown as filled squares. If no pixels in that bin are above the $3\sigma$ significance level, we report the 2$\sigma$ surface brightness upper limit as non-detection (open squares). Figure \ref{fig:source_radial_OII_FeII_MgII}, left panel, shows that the \oii\ surface brightness radial profile extends out to $\approx 15.0_{-0.3}^{+0.4}$ kpc.

As gravitational lensing allows us to zoom-in on smaller regions around the highly magnified image of the main arc (Figure \ref{fig:muse_hst}), we can probe and test if outflow emission is spatially extended around individual star-forming regions within the galaxy. We use images 1, 2, and 3 (Figure \ref{fig:muse_hst}) to trace the {\oii}, {\feii}, and \mgii\ emission similar to what is done for the counter arc. We individually reconstruct the \oii\ emission maps around three multiply lensed regions of the main arc (see Figure 1). Figure \ref{fig:sourceOII_FeII_MgII_full_3EUB}, top row, shows the source-plane reconstructed {\oii} emission maps around these regions. While Figure \ref{fig:sourceOII_FeII_MgII_full_3EUB} spatially resolves many of the bright star-forming regions, regions E and B are blended in image 1 (left column). Image 2 has the highest spatial resolution in the source plane as it is highly magnified compared to the other images, such that the regions U, E, and B are spatially resolved (middle column). From these images, we measure the radial profiles of nebular emission for the individual regions U, E, and B, in addition to the full scale of the galaxy from the more distorted image 3. Figure \ref{fig:src_radial_3EUB_OII_FeII_MgII} shows the {\oii} radial profiles for these regions. The {\oii} emission maximum radial extents are $\approx 20.5_{-1.0}^{+1.0}$ kpc in image 3, $5.7_{-0.2}^{+0.3}$ kpc in region U, $8_{-0.2}^{+0.2}$ kpc in region E, and $6.7_{-0.1}^{+0.1}$ kpc in region B. However, image 3 is the most distorted and has light contamination from two foreground cluster galaxies \cite[See Figure 1 in][]{wuyts2014magnified}. Therefore, we take the radial distances measured from the counter arc as a more reliable estimate of the radial extent of the whole galaxy, and include image 3 for completeness.

The {\oii} emission surface brightness above $3\sigma$ significance in the counter arc image covers an area of $115_{-10}^{+4} \mathrm{kpc^2}$ in the source plane of the galaxy. We compute this area by computing the total area of all pixels exhibiting $>3\sigma$ significant \oii\ emission. We sample 100 realizations of the lens model to quantify the uncertainties on the measured area. The total observed area of the entire counter arc in Figure \ref{fig:source_OII_FeII_MgII_full_CARC} that covers a radial distance of 30 kpc is ${\rm 1387_{-76}^{+43}\ kpc^2}$. This means that the fraction of the total area covered by the {\oii} emission, that traces the nebular regions in the galaxy, is $8.29_{-0.85}^{+0.39}\ \%$.

\subsection{\feii\ Emission}
We further study the spatial extent of the prominent \feii\ fluorescent or non-resonant emission (see Figure \ref{fig:muse_FeII_lines}). \feii\ fluorescent emission in outflowing gas arises owing to the de-excitation of the resonant Fe II absorption lines. The photons are re-emitted at different wavelengths than those of the absorption lines. This happens because the electrons move from the excited state to one of the ground state levels close to the original ground state level but with slightly different energy due to the fine-structure splitting of the ground state \citep{prochaska2011simple}.
To maximize the SNR, we construct one combined \feii\ emission map by combining five narrow emission regions traced by \feii\ emission lines at 2365{\AA}, 2396{\AA}, 2612{\AA}, 2626{\AA}, and 2632 {\AA} as described in Section \ref{2D_maps}.

\subsubsection{\feii\ Emission in the image plane}
We see a statistically significant ($\geq$ 3$\sigma$) {\feii} (continuum-subtracted) emission in the image plane for both the main arc and the counter arc in the upper and lower panels of the middle column of Figure \ref{fig:muse_OII_FeII_MgII_map_full}, respectively. There is significant {\feii} emission in the image plane in both cases. As \feii\ fine-structure emission lines may trace the most dense regions of the outflowing gas \citep{prochaska2011simple}, this result suggests that the most dense part of the outflowing gas may reside relatively close to the star-forming regions of this galaxy.

\subsubsection{\feii\ emission in the source plane}
We reconstruct the continuum-subtracted {\feii} emission maps in the source plane as described in Section \ref{sec:reconstruction}. Figure \ref{fig:source_OII_FeII_MgII_full_CARC}, middle panel, shows the source plane reconstruction of the statistically significant ($\geq$ 3$\sigma$) {\feii} emission map around the counter arc. The 3$\sigma$ emission contours (bottom panel) overlaid on the source-plane reconstructed HST image are significantly more extended than the starlight being traced by the HST image. Figure \ref{fig:spatial_XY_extent_src} (middle panels) show the observed maximum spatial extent along the x-axis and y-axis of \feii\ emission. After correcting for the lensing shear+seeing, the maximum extent along the x-axis is $ 21.0_{-0.4}^{+0.4}$ kpc, and the maximum extent along the y-axis is $ 13.7_{-0.3}^{+0.4}$ kpc, respectively. These measurements are summarized in Table \ref{tab:CARC_MgII_distances}. Furthermore, \feii\ emission is more spatially extended than the nebular \oii\ emission. The {\feii} emission radial profile reaches an observed radial distance of $14.33_{-0.25}^{+0.27}$ kpc from the center of the galaxy in the counter arc. This is shown in the middle panel of Figure \ref{fig:source_radial_OII_FeII_MgII}. 

The middle row of Figure \ref{fig:sourceOII_FeII_MgII_full_3EUB} shows the source plane reconstructions of statistically significant ($\geq$ 3$\sigma$) \feii\ emission in images 1, 2, and 3, respectively. The \feii\ emission in these images is more spatially extended compared to the corresponding \oii\ emission (top row). This further shows that the \feii\ emission tracing the cool outflow is extended beyond star forming regions in all the three images in the source plane. The purple points in Figure \ref{fig:src_radial_3EUB_OII_FeII_MgII} show the {\feii} surface brightness radial profiles around the three star-forming regions E, U, and B. The {\feii} emission profile is more spatially extended relative to the nebular {\oii} emission in all regions. This is further evidence that the outflow traced by the {\feii} emission is more spatially extended than the star-forming regions in the galaxy. The {\feii} radial profiles extend out to $26.5_{-1.4}^{+1.5}$ kpc in image 3, $9_{-0.2}^{+0.4}$ kpc in region U, $9.2_{-0.2}^{+0.2}$ kpc in region E, and $10_{-0.1}^{+0.1}$ kpc in region B.

We also measure the area covered by the {\feii} emission above $3\sigma$ significance in the counter arc image as $298_{-14}^{+9} \mathrm{kpc^2}$. This corresponds to $21.49_{-1.55}^{+0.93}\%$ of the total area of the counter arc stamp in the source plane. This area is almost $\sim 2.6$ times the area covered by {\oii} emission. 

\subsection{\mgii\ Emission}
We investigate the spatial extent of the \mgii\ emission doublet around this galaxy. We construct the stellar continuum and \mgii\ emission+continuum surface brightness maps to produce the (continuum-subtracted) \mgii\ emission maps as described in Section \ref{2D_maps}. 

\subsubsection{\mgii\ Emission in the image plane}\label{sec:MgII_ems_img_plane}
\begin{figure*}
    \centering
    \includegraphics[width=\textwidth]{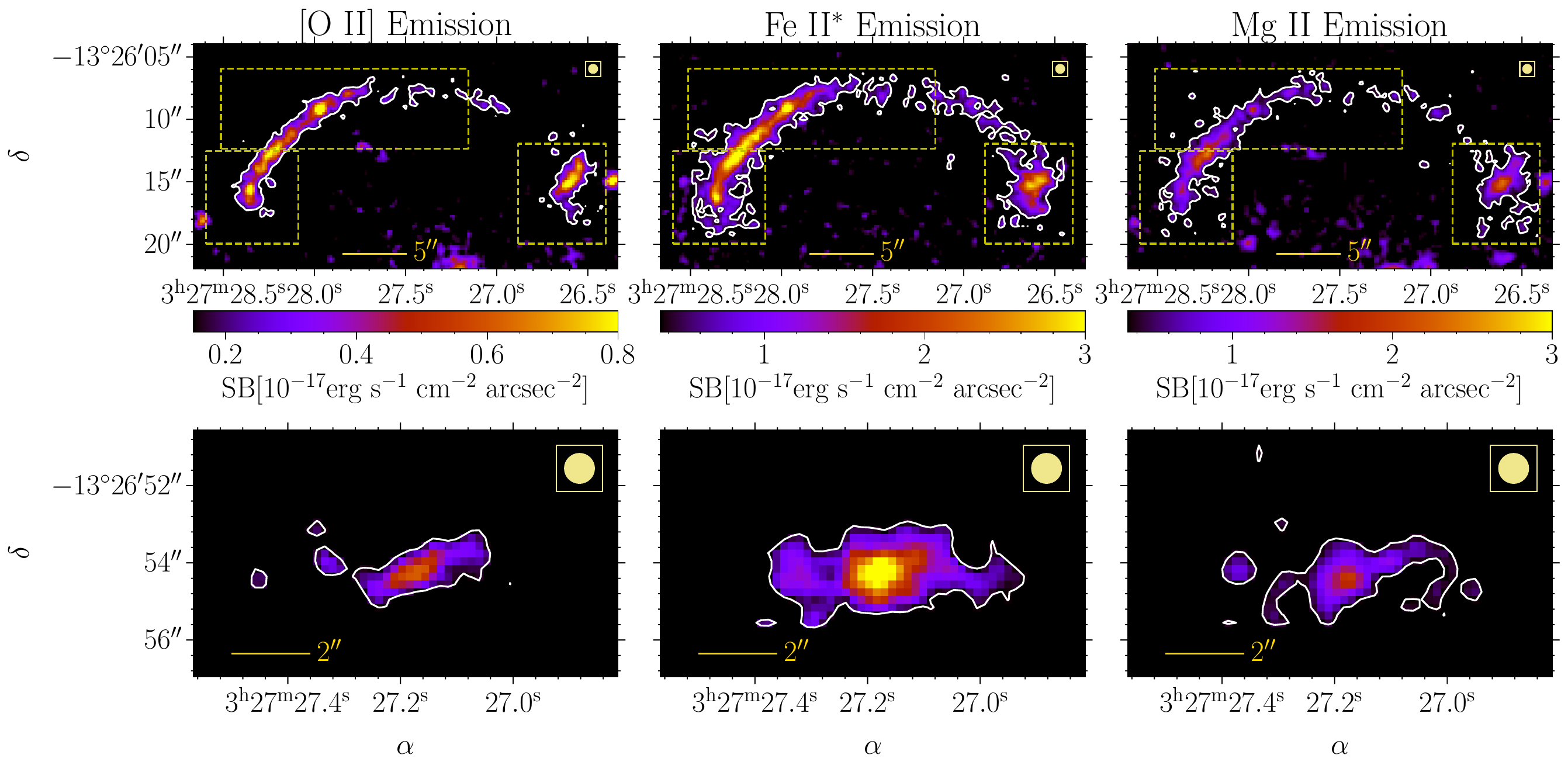}
    \caption{ Surface brightness maps in the image plane for the continuum-subtracted {\oii} emission (\textit{Left}), {\feii} emission (\textit{Middle}), and {\mgii} emission (\textit{right}) around the main arc (\textit{top row}) and the counter arc (\textit{bottom row}), respectively. The \textit{white} contours correspond to 3$\sigma$ surface brightness significance levels (Table \ref{tab:sigma_values}). The yellow dashed rectangles in the top row correspond to the three images of the main arc from Figure \ref{fig:muse_hst} from left to right, respectively. The yellow circles in the top right of each subplot represent the maximum seeing during observation.}
    \label{fig:muse_OII_FeII_MgII_map_full}
\end{figure*}

To maximize the SNR, we first create the integrated \mgii\ emission+continuum maps by summing both the 2796 {\AA} and 2803 {\AA} emission doublets together (see Figure \ref{fig:stack_spec_mgII_emission}). We refer to them as the full {\mgii} emission maps. Figure \ref{fig:muse_OII_FeII_MgII_map_full} right panels show the image plane map of continuum-subtracted \mgii\ emission for the main arc (top panel) and the counter arc (bottom panel), respectively.
The white contours represent the $3\sigma$ significance level {\mgii} surface brightness. In the main arc (top right panel), \mgii\ emission in the image plane is mostly extended around image 3 and region U, and has the least spatial extent around region B. \mgii\ emission is also spatially extended around the less magnified counter arc. In all cases, we detect a significant (continuum-subtracted) \mgii\ emission. 

\begin{figure*}
    \centering
    \includegraphics[width=\textwidth]{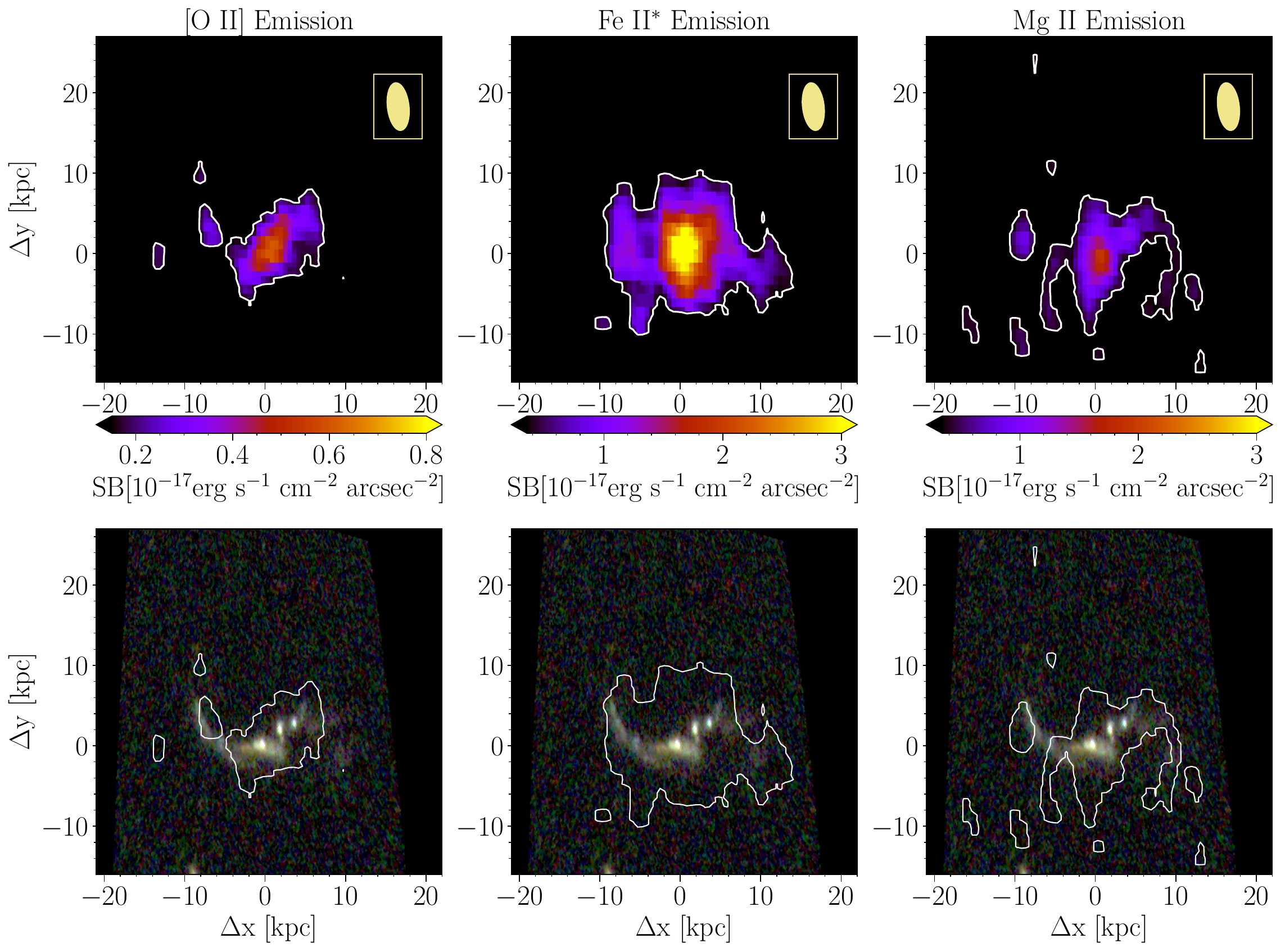}
    \caption{Source plane reconstructions of the {\oii} emission, {\feii} emission, and the  \mgii\ emission doublet surface brightness map around the counter arc. \textit{Top Row:} The reconstructions of the MUSE surface brightness maps for the emission lines in the source plane. The white contours represent the 3$\sigma$ significance levels (see Table: \ref{tab:sigma_values}). The ${\rm \Delta x}$, and ${\rm \Delta y}$ represent the distance along the right ascension $\alpha$ direction, and declination $\delta$ direction, respectively. The yellow ellipses in the top panels represent the maximum seeing in the source plane for the counter arc. \textit{Bottom Row:} Source plane reconstruction of the HST image for the counter arc. We plot the 3$\sigma$ contours from the top row over these HST reconstructions.} 
    \label{fig:source_OII_FeII_MgII_full_CARC}
\end{figure*}

\begin{figure*}
    \centering
    \includegraphics[width=\textwidth]{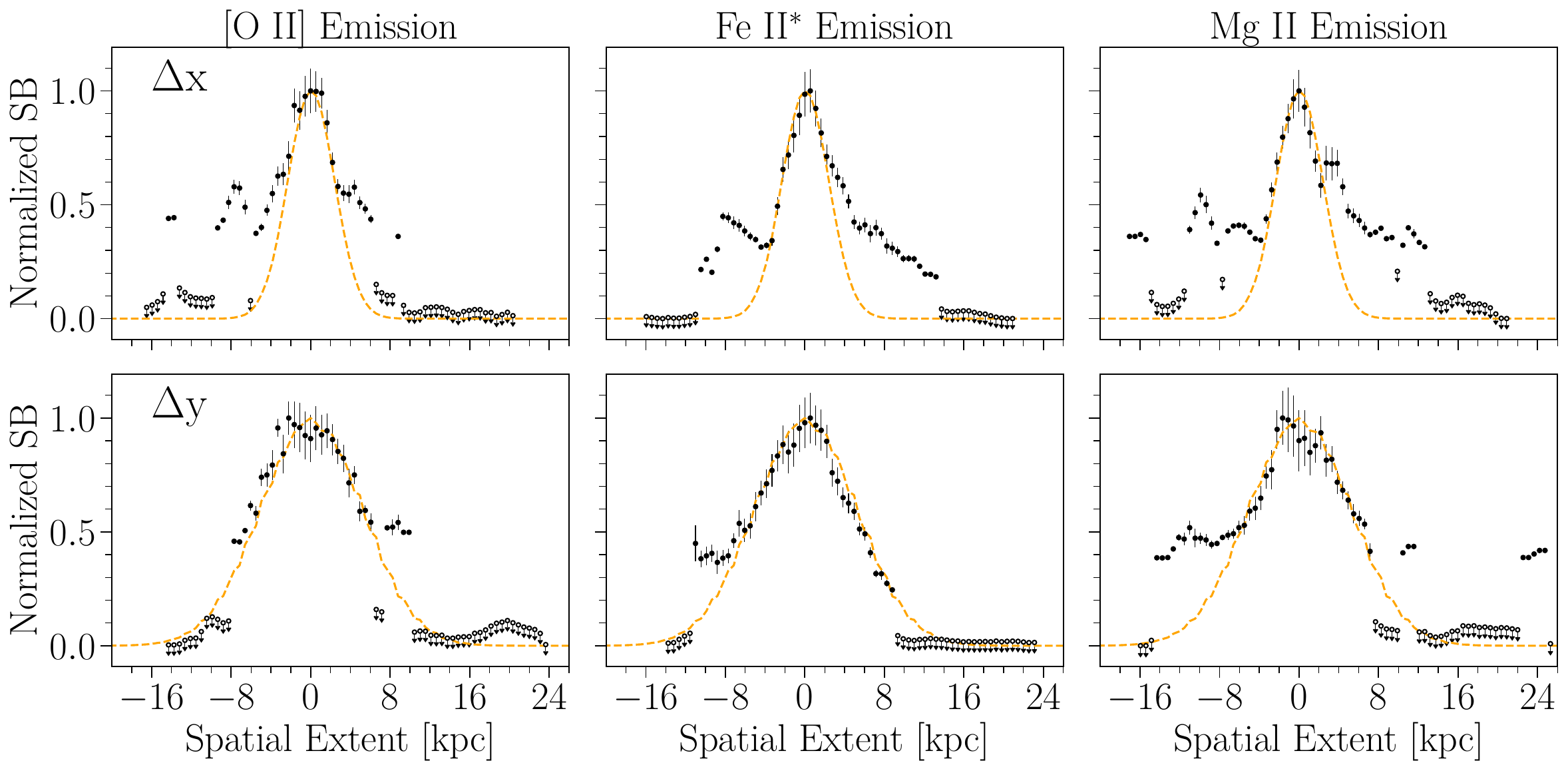}
    \caption{Source plane surface brightness spatial extents of the {\oii} emission (\textit{left}), {\feii} emission (\textit{middle}), and the \mgii\ full emission (\textit{right}) in the counter arc. The \textit{top row} and \textit{bottom row} represent the spatial extent of the surface brightness along the x- (right ascension $\alpha$) and y-direction (declination $\delta$), respectively. The open circles with arrows represent the 2$\sigma$ upper limit for non-detections. All the surface brightness data points are normalized by dividing by the value of the pixel with maximum surface brightness. The yellow dashed lines in the top and bottom rows represent the normalized seeing profile extents along the x- and y-directions, respectively.}
    \label{fig:spatial_XY_extent_src}
\end{figure*}

\begin{figure*}
    \centering
    \includegraphics[width=\textwidth]{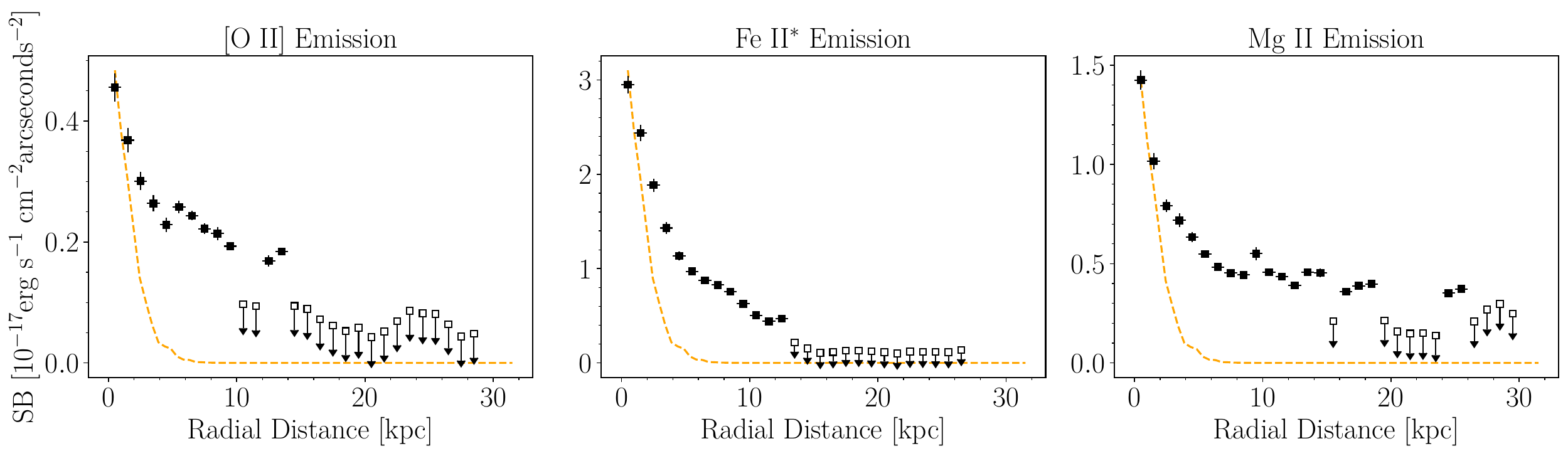}
    \caption{Surface brightness radial profiles for the {\oii} emission (\textit{Left}), {\feii} emission (\textit{middle}), and {\mgii} emission (\textit{right}). The radial profiles are measured from the center of the galaxy. The filled \textit{black} squares represent radial bins with significance greater than 3$\sigma$. The open \textit{black} squares represent the 2$\sigma$ upper limits in radial bins with non-detections. The \textit{dashed gold} line represents the radial profile for the seeing for the counter arc image in the source plane.}
    \label{fig:source_radial_OII_FeII_MgII}
\end{figure*}

\begin{figure*}
    \centering
    \includegraphics[width=\textwidth]{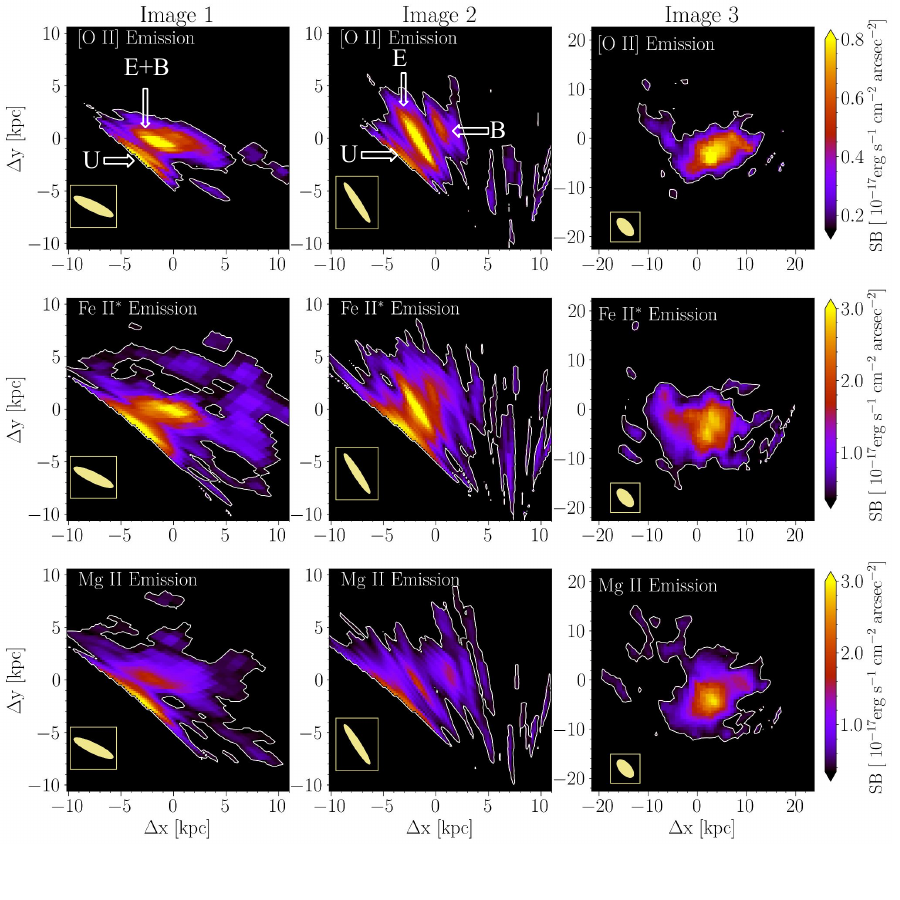}
    \caption{Source plane reconstruction of the {\oii} emission (\textit{top row}), the {\feii} emission (\textit{middle row}), and the \mgii\ emission doublet $\lambda \lambda$ 2796, 2803 {\AA} emission (\textit{bottom row}) surface brightness maps in images 1 (\textit{left column}), 2 (\textit{middle column}) and 3 (\textit{right column}), respectively. The \textit{white} contours represent the 3$\sigma$ significance levels. We label the four individual star-forming regions as shown in Figure \ref{fig:muse_hst}. The reconstructed maximum seeing is shown as a filled \textit{yellow} region in the left corner of each subplot.} 
    \label{fig:sourceOII_FeII_MgII_full_3EUB}
\end{figure*}

\begin{figure*}
    \centering
    \includegraphics[width=\textwidth]{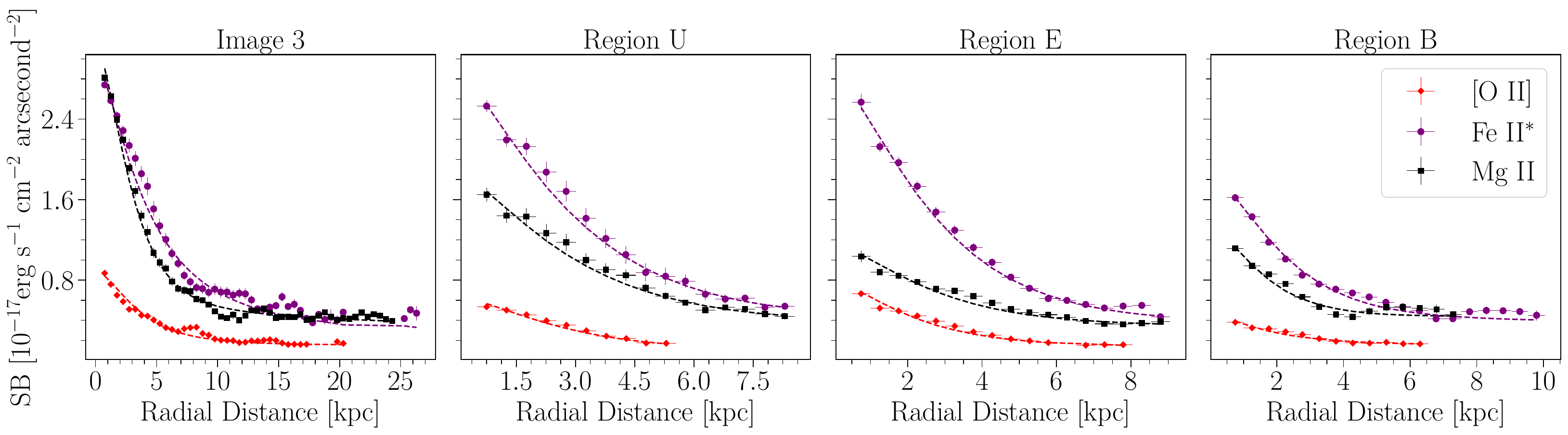}
    \caption{ Mean surface brightness radial profiles for the {\oii} (\textit{red diamonds}), {\feii} (\textit{purple circles}), and {\mgii} emission (\textit{black squares}) in the source plane for image 3 (\textit{first panel}), the star-forming regions U (\textit{second panel}), E (\textit{third panel}), and B (\textit{fourth panel}). All the data points are representatives of radial bins with surface brightness pixels greater than 3$\sigma$. The dashed lines represent the fitted models from Section \ref{sec:Method_Radial_Profiles}. The best fit parameters are summarized in Table \ref{table:radial_parameters}.1 in Appendix \ref{sec:appendix_radial_and_Cf}. The {\feii}, and {\mgii} are bright and more extended than the {\oii} nebular emission. }
    \label{fig:src_radial_3EUB_OII_FeII_MgII}
\end{figure*}

The \mgii\ 2803 {\AA} emission spectrum shows a unique feature of two kinematically distinct emission components (Figure \ref{fig:stack_spec_mgII_emission}). Both these components are redshifted relative to the systemic redshift of the host galaxy. The stronger emission peak is observed at a mean velocity of 100 ${\rm km\ s^{-1}}$ and we classify it as the primary peak. The center of the weaker emission component is redshifted by 500 ${\rm km\ s^{-1}}$ from the systemic redshift of the galaxy, and we call it the secondary peak (marked with green band in Figure \ref{fig:stack_spec_mgII_emission}, right panel). The average SNR per pixel for this component is 78 and 19 in the light-weighted spectra of the main arc and the counter arc, respectively. It is very rare to find two distinct \mgii\ emission components separated by $\Delta v \approx 400\ {\rm km\ s^{-1}}$ and this suggests that they may arise from two distinct past outflow events. We investigate if these unique features are co-spatial or originate in different parts of the galaxy. We construct two narrowband \mgii\ emission maps around the primary and the secondary peaks as described in Section \ref{2D_maps}. 

\begin{figure*}
    \centering
    \includegraphics[width=\textwidth]{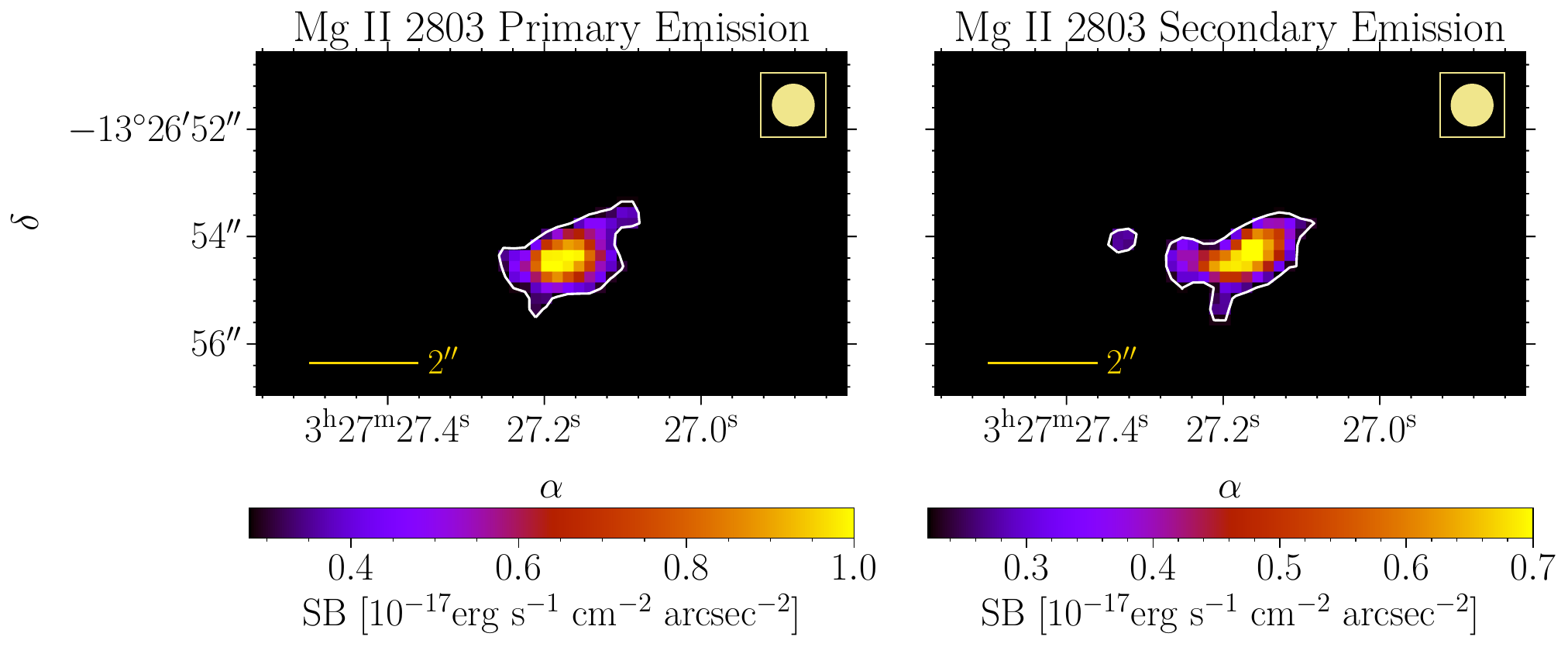}
    \caption{{\mgii} continuum-subtracted emission surface brightness maps for the Mg II 2803 primary (\textit{Left}) and secondary (\textit{Right}) components, respectively. The \textit{white} contours represent the 3$\sigma$ surface brightness limits.} The \textit{yellow} circles in the top left of each subplot represent the maximum seeing during observation. Both the primary and secondary \mgii\ emission peaks are spatially extended. The primary component and the secondary component are redshifted and have mean velocities of 100 ${\rm km\ s^{-1} }$ and 500 ${\rm km\ s^{-1}}$ with respect to the galaxy redshift, respectively.
    \label{fig:muse_mgII2803_primary_secondary_CARC}
\end{figure*}

Figure \ref{fig:muse_mgII2803_primary_secondary_CARC} shows the image plane surface brightness maps of the \mgii\ 2803 {\AA} primary (left panel) and secondary (right panel) emission, constructed around the counter arc. The white contours represent the 3$\sigma$ significance level for both lines. The morphological difference between the two emission maps suggests that there are two outflowing components with two different velocities. Both of these components show spatially extended \mgii\ emission.

\subsubsection{\mgii\ Emission in the source plane}\label{sec:MgII_ems_src_plane}
To quantify the spatial extent of the \mgii\ emission, we transform the narrowband images into the source plane of the galaxy. We first reconstruct the \mgii\ (continuum-subtracted) emission maps around the counter arc, as described in Section \ref{sec:reconstruction}. Figure \ref{fig:source_OII_FeII_MgII_full_CARC}, top right panel, shows the reconstructed \mgii\ emission map in the galaxy's source plane for the counter arc. The white contours represent statistically significant ($\geq $ 3$\sigma$) surface brightness emission. These 3$\sigma$ contours are shown in the source plane reconstructed HST image of the counter arc (Figure \ref{fig:source_OII_FeII_MgII_full_CARC}, bottom right panel). The {\mgii} emission has a different morphology compared to the shape of the galaxy in this HST image. Furthermore, Figure \ref{fig:source_OII_FeII_MgII_full_CARC} shows that the {\mgii} emission is more extended compared to the nebular {\oii} emission (top left panel).

The \mgii\ emission is clearly asymmetric and shows small-scale structure. Individual bright \mgii\ emission knots are detected between 10-20 kpc from the center of the galaxy at different incidence. Further, the total observed projected area covered by outflowing gas is ${\rm 184_{-10}^{+5}\ kpc^{2}}$. This area is $\sim$ 1.6 times the area covered by the {\oii} emission in the source plane. This exercise clearly demonstrates the clumpy nature of the outflow. We estimate that $13.27_{-1.02}^{+0.55}$\% of the area of the field of view in Figure \ref{fig:source_OII_FeII_MgII_full_CARC} shows statistically significant \mgii\ emission.

The \mgii\ emission is clearly spatially asymmetric in Figure \ref{fig:source_OII_FeII_MgII_full_CARC}. The observed surface brightness profile is more extended along the declination direction (${\rm \Delta \delta\ or\ \Delta y =38.9_{-0.6}^{+0.8}}$ kpc) than along the right ascension direction (${\rm \Delta \alpha\ or\ \Delta x = 30.5_{-0.6}^{+0.6}}$ kpc), exhibiting $\sim 27.6_{-0.7}^{+0.8}$\% more extent along the y-direction than along the x-direction at 3$\sigma$ significance. The x (upper panel) and y (lower panel) extents of the {\mgii} surface brightness profiles of the counter arc are shown in the right column of Figure \ref{fig:spatial_XY_extent_src}.

To estimate the true spatial extent, we correct for seeing and lensing shear effects as described in Section \ref{sec:shearing_lensing_correction}, and quantify the maximum spatial extent of \mgii\ emission in this galaxy as ${\rm \Delta \delta\ or\ \Delta y = 30.0_{-0.5}^{+0.7} }$ kpc and ${\rm \Delta \alpha\ or\ \Delta x = 24.8_{-0.5}^{+0.5} }$ kpc, respectively. Even after correcting for atmospheric seeing and lensing shear, the \mgii\ emission is spatially more extended along the declination direction by $20.9_{-0.6}^{+0.7}$\% (See Table \ref{tab:CARC_MgII_distances}).

The \mgii\ emission has some extended shell-like structure at $\sim$ -5 and 10 kpc in the $\Delta x$ direction (Figure \ref{fig:source_OII_FeII_MgII_full_CARC}, top right panel). These shells could be associated with a spiral arm of the galaxy itself. To test this hypothesis, we extract three light weighted spectra: One for the left shell, one for the right shell, and one for the middle core. For these three regions, we compute the systemic redshift by using the \oii\ emission line and compute the kinematics of the \mgii\ emission components. The detailed models will be described in a future paper (Shaban et al. in preparation) but the resultant \mgii\ emission velocities are almost identical within error bars for all three regions, whereas the systemic redshifts are varying. This rules out the possibility that these structures are associated with the spiral arms of the galaxy, and are indeed gas structures only traced by \mgii\ emission. This is also seen qualitatively in the bottom right panel of Figure \ref{fig:source_OII_FeII_MgII_full_CARC}. The source-plane reconstructed HST image traces the stars of the galaxy, which is spatially offset from the location of the \mgii\ emitting regions.

In Appendix \ref{sec:MgII_components_extent}, Figures \ref{fig:MgII2803I_radial_x_y} and \ref{fig:MgII2803II_radial_x_y} show the source plane reconstruction of the counter arc showing the \mgii\ 2803 primary and \mgii\ 2803 secondary emission components surface brightness maps, respectively. The \mgii\ 2803 primary and secondary components are kinematically separated by $\Delta v \approx 400\ {\rm km\ s^{-1}}$ and the secondary component is detected in almost all regions of the galaxy except in region U. In the source plane, we detect a significant \mgii\ emission ($> 3\sigma$) for both the primary and secondary \mgii\ 2803 {\AA} emission peaks. The two components have similar projected spatial extent, but are asymmetric. After correcting for seeing and lensing shear, the \mgii\ 2803 {\AA} primary emission component is more spatially extended than the secondary one along the declination direction by $\approx$ 2.5 kpc. For the \mgii\ 2803 {\AA} secondary emission component profile is more extended along the right ascension direction by $\approx$ 2.5 kpc, after seeing+shear correction. The observed \mgii\ emission surface brightness profiles are presented in Appendix \ref{sec:MgII_components_extent} and the distance measurements are summarized in Table \ref{tab:CARC_MgII_distances}.

As these two components are kinematically distinct, they may exist in different physical regions along the line of sight. This can be interpreted as the bulk of the two \mgii\ emitting components being at different velocities. There may be some kinematic overlap between the two components, as they appear partially blended in the 1D spectrum (Figure \ref{fig:stack_spec_mgII_emission}). Their kinematic offset combined with different morphology of the two emission lines suggest that they may have different origin (e.g. different star-bursts, or originating from different star-forming regions), although they appear approximately co-spatial in projection. Further analysis of the absorption lines will provide more insight regarding the line of sight geometry of the outflow.

\begin{deluxetable*}{ccccc}
\tablecolumns{5}
\tablewidth{0pt}
\tablecaption{Maximum spatial extent of the {\oii}, {\feii}, and {\mgii} emission in the source plane using the counter arc}.
\tablehead{
\colhead{Transition/Component} &
\colhead{Observed $\Delta x$ [kpc]} &
\colhead{Observed $\Delta y$ [kpc]} &
\colhead{Corrected $\Delta x$ [kpc]\tablenotemark{a}} &
\colhead{Corrected $\Delta y$ [kpc]\tablenotemark{a}}}
\startdata
{\oii} Emission & $24.1_{-0.5 }^{+0.5}$ & $17.3_{-0.3}^{+0.4}$ & $13.5 _{-0.2 }^{+0.3}$ & $9.2_{-0.3}^{+0.4}$ \\
{\feii} Emission & $26.1_{-0.5}^{+0.5}$ & $21.6_{-0.3}^{+ 0.4}$ & $21.0_{-0.4}^{+0.4}$ & $13.7_{-0.3}^{+ 0.4}$\\
Full {\mgii} Emission & $30.5 _{-0.6}^{+ 0.6}$ & $38.9 _{-0.6}^{+0.8}$ & $24.8 _{-0.5}^{+0.5}$ & $30.0_{-0.5}^{+0.7}$ \\
Primary {\mgii} 2803 {\AA} Component & $8.9 _{-0.2}^{+ 0.2}$ & $17.0_{-0.2}^{+0.3}$ & $6.4_{-0.2}^{+0.2}$ & $8.9_{-0.2}^{+0.3}$ \\
Secondary {\mgii} 2803 {\AA} Component & $13.7_{-0.3}^{+0.3}$ & $16.1_{-0.2}^{+0.3}$ & $10.5 _{-0.2}^{+0.2}$ & $ 8.0_{-0.2}^{+0.2}$ \\
\enddata
\vspace{-0.2cm}
\label{tab:CARC_MgII_distances}
\tablenotetext{a}{After seeing and lensing shear correction (see Section \ref{sec:MgII_ems_src_plane}).}
\end{deluxetable*}

\subsubsection{\mgii\ Emission Radial Profiles}
We characterize the continuum subtracted mean surface brightness profile as a mean 1D residual Mg II emission profile in Figure \ref{fig:source_radial_OII_FeII_MgII}, right panel. We define the center of the radial profile as the brightest pixel of the counter-arc image. In each radial bin, we compute the mean surface brightness of all pixels above the 3$\sigma$ background level, shown as filled squares. If no pixels in that bin are above the 3$\sigma$ background level, we report the 2$\sigma$ background surface brightness level for non-detection (open squares). patchy \mgii\ emission is detected out to an observed radial distance of $26.5_{-0.4}^{+0.5}$ kpc. For comparison, the dashed golden line shows the maximum seeing as present in the reconstructed source-plane. Clearly, \mgii\ emission is spatially extended.

Figure \ref{fig:sourceOII_FeII_MgII_full_3EUB}, bottom row, shows the reconstructed \mgii\ emission maps in the source plane for the three images. From all the source plane reconstructed images, the \mgii\ emission contours extend beyond the {\oii} emission contours at the $3\sigma$ significance level. The \mgii\ emission (Figure \ref{fig:sourceOII_FeII_MgII_full_3EUB}, bottom panels) is significantly extended spatially, with emission arising both in the bright star-forming clusters (e.g., E, B, and U) as well as a diffuse spatially extended component.

We extract the mean surface brightness radial profiles for the \mgii\ emission as described in \ref{sec:Method_Radial_Profiles}. Figure \ref{fig:src_radial_3EUB_OII_FeII_MgII} shows these radial profiles for these regions as solid black points. Figure \ref{fig:src_radial_3EUB_OII_FeII_MgII} shows that the \mgii\ emission extends to $\approx$ $9.6_{- 0.2 }^{+0.2}$ kpc, $9.0_{-0.2}^{+0.3}$ kpc, and $7.4_{-0.1}^{+0.1}$ kpc in regions E, U, and B, respectively, as measured from the brightest central pixel of each individual region. The \mgii\ emission in image 3 extends radially up to $24.5_{-1.6}^{+1.6}$ kpc. We find that the 3$\sigma$ maximum radial extents of the \mgii\ emission are $\sim$ 25 kpc for both image 3 and the counter arc. Image 3 and the counter arc provide the extent of the \mgii\ emission in the galaxy as a whole.
However, image 3 is much more magnified and distorted compared to the counter arc. Furthermore, image 3 has some contribution from two foreground cluster galaxies \cite[See Figure 1 in][]{wuyts2014magnified}. Therefore, the counter arc provides a more reliable measure for the radial extent of the outflow traced by the {\mgii} emission.

\subsection{Covering fraction}
\begin{figure}
    \centering
    \includegraphics[width=0.5\textwidth]{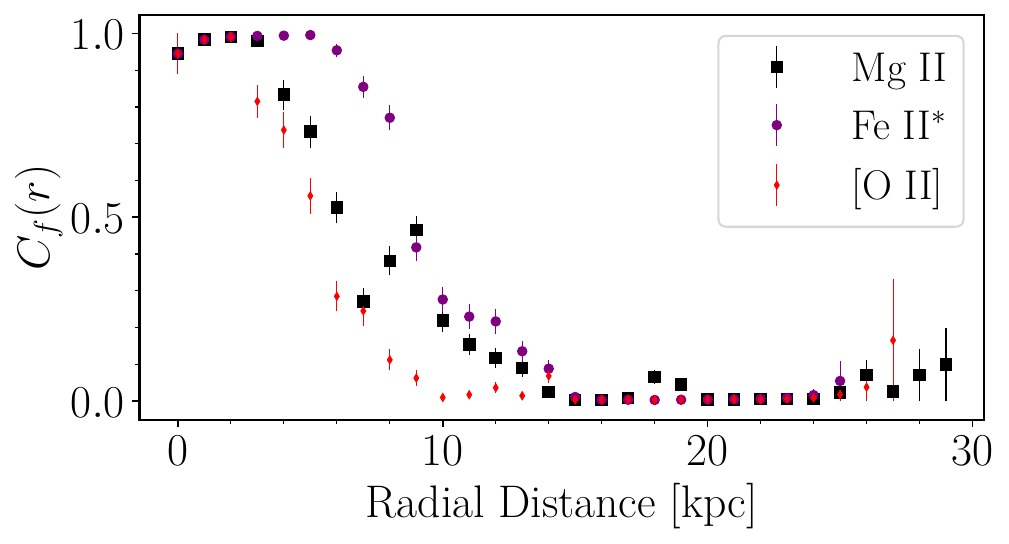}
    \caption{Gas covering fraction $C_f(r)$ as a function of distance for \mgii\ emission (\textit{black squares}), {\feii} emission (\textit{purple circles}), and {\oii} emission (\textit{red diamonds}), respectively. The detection limit is $3\sigma$ for the surface brightness of a pixel to be counted as a detection. A value of $C_f(r) = 1$ means the radial bin area is totally covered by emission, and a value of $C_f(r) = 0$ means that there is no emission covering any area of that radial bin.} 
    \label{fig:covering_fraction}
\end{figure}

The covering fraction in this work represents the fraction of area around the galaxy which is covered by the continuum-subtracted emission (in radial bins), traced by the {\oii}, {\feii}, and {\mgii} above the $3\sigma$ limit (see Section \ref{sec:covering_fraction} for details). We choose the brightest pixel of the reconstructed counter arc image as the center (Figure \ref{fig:source_OII_FeII_MgII_full_CARC}, top row). We compute the fraction of pixels in each radial bin that are above the $3\sigma$ limit, out to 30 kpc. We select this distance limit to avoid any contributions from other bright foreground objects near the counter arc. The covering fraction can be interpreted as a measure of the porosity or patchiness of the outflowing gas \citep{martin2009physical,chisholm2016robust,chisholm2018feeding}. In this work, we use a different approach to measure $C_f(r)$. We measure the total area (pixels) exhibiting statistically significant emission to calculate $C_f(r)$. We calculate $C_f(r)$ in the source plane reconstructed counter arc only because it represents the full galaxy.
Figure \ref{fig:covering_fraction} shows the measured covering fraction as function of radial distance for the continuum-subtracted {\oii}, {\feii}, and \mgii\ emission, respectively. The {\oii} $C_f$ measurements (red diamonds), are almost unity within 2 kpc and then fall off sharply, reaching zero for all bins beyond 10 kpc.

{\mgii} $C_f$ is nearly unity in the inner 3 kpc, which means that the outflowing gas is ubiquitous and totally covers the area within these inner radii. As we go outward from 3 kpc to 10 kpc, the covering fraction drops to $\sim 20\%$. From 10 kpc to 30 kpc, it gradually decreases and oscillates between 0 - and 10 \%. These fluctuations indicate that the outflowing gas is not uniformly distributed and is patchy. There are regions, where larger concentrations of outflowing gas exist even at large radial distances, and there are regions where little outflowing gas is detected. This reaffirms the canonical picture of a large-scale patchy galactic outflow that is being traced by the \mgii\ emitting gas.

The {\feii} emission maps have much higher SNR compared to the {\mgii} ones, as they are constructed by adding five distinct emission lines. The {\feii} $C_f$ is unity within the first 5 kpc (purple circles), after which it falls off radially, and almost reaches zero at 15 kpc. In all inner radial bins ($<$15 kpc), {\feii} exhibits higher $C_f$ than {\mgii}, however, only {\mgii} emission exhibit non-zero $C_f$ at higher radii.

Both the {\feii} and {\mgii} $C_f$ measurements are higher than that of {\oii} emission. At each radial bin, there are more pixels covered by significant {\feii} and {\mgii} emission than {\oii} nebular emission. This provides further evidence that the galactic outflow traced by the {\feii} and {\mgii} emission are more spatially extended compared to the stellar nebular emission. 
We characterize the \oii, \feii, and \mgii\ emission covering fraction radial profiles using a power law:
\begin{equation}
    C_f (r) = C_{f,0} \left(\frac{r}{1\mathrm{kpc}}\right)^{\gamma}\label{eq:cf}
\end{equation}
where $C_{f,0}$ is the covering fraction at the center of the galaxy, and $\gamma$ is the power law index. We convolve this power law with the seeing of the counter arc in the source plane. 

We obtain the best fit model for {\oii} with $\gamma = -1.63_{-0.03}^{+0.03}$, {\mgii} with $\gamma= - 1.25_{-0.02}^{+0.02}$, and {\feii} with $\gamma = -1.16_{-0.01}^{+0.01}$. The best fit models for $C_f$ of the three emission transitions are shown in Figure \ref{fig:Cf_modeling_OII_FeII_MgII} in Appendix B. We constrain the total area covered by the emission tracing the outflow and the total area enclosed within 30 kpc. This 30 kpc radial distance limit marks the boundary of the observed area of the counter arc in the source plane in Figure \ref{fig:source_OII_FeII_MgII_full_CARC}. By dividing the outflow area $A_{out}$ by the total area $A_{tot}$ of the counter arc stamp, we can get an average value for the covering fraction. The measured value for the average covering fraction is $\langle C_f \rangle = \frac{A_{out}}{A_{tot}} = 0.0829_{-0.009}^{+0.004}$, $0.21_{-0.02}^{+0.01}$, and $0.13_{-0.01}^{+0.01}$ for the {\oii}, {\feii}, and {\mgii}, respectively.

The covering fraction beyond the stellar continuum is an indicator of the morphology and patchiness of the outflowing gas. In other words, it quantifies the fraction of the projected area around a galaxy where outflowing gas can be detected \citep{chisholm2016robust}. Several studies quantified the gas covering fraction using partial covering of blueshifted absorption lines, with some assumption about the relation between the velocity and radius \cite[e.g.,][]{chisholm2016robust, chisholm2018feeding}. Typically, these works find a decreasing $C_f$ with distance characterized by a power-law. Our measurement of the {\mgii} $C_f$ power law is comparable to these studies, even though a completely different approach is being used here. Additional analysis of these two methods is needed to compare if the \mgii\ emission traced gas covering fraction, and the absorption traced line-of-sight covering fraction are indeed probing the same gas covering fraction. This will be done in \rcsa\ as a part of a future paper (Shaban et al. in prep). Covering fraction is one of the important quantities in the calculation of the mass outflow rate $\dot{M}_{out}$. Most studies assume $C_f$ to be constant. Our measurements conclusively show that the outflow gas covering fraction changes as we move outward from the central region of the galaxy. These constraints will enable robust mass outflow rates for this system. 

\section{Discussion and Conclusions}\label{sec:discuss}
In this paper we present observations of  \oii, \feii, and \mgii\ emission lines in \rcsa, a $z=1.7034$ galaxy, lensed by a foreground galaxy cluster at $z\approx0.56$. Lensing distorts light coming from the background ``source" galaxy that results in multiple images of the source galaxy \rcsa\ at the plane of the cluster, referred to as the ``image plane". Some of these images are highly magnified and represent small star-forming regions within the galaxy (e.g. images 1 and 2 in \autoref{fig:muse_hst}). Other images are less magnified and represent the full extent of the galaxy (e.g. image 3 and the counter arc in \autoref{fig:muse_hst}). With VLT/MUSE observations of these different images, we study the spatial extent of the emission lines tracing galactic outflows, and provide strong constraints on the spatial extent of the outflowing gas. Our main results are:

\begin{itemize}
    \item We detect and compare the \oii\ nebular emission with the stellar continuum around the galaxy. The \oii\ nebular emission is not spatially extended compared to the stellar continuum (Figures \ref{fig:muse_OII_FeII_MgII_map_full} \& \ref{fig:source_OII_FeII_MgII_full_CARC}). The \oii\ emission shows maximum spatial extent along the x- and y-directions after correcting for shear lensing+seeing of $13.5_{-0.2}^{+0.3}$ kpc, and $9.2_{-0.3}^{+0.4}$ kpc, respectively (Figure \ref{fig:spatial_XY_extent_src} \& Table \ref{tab:CARC_MgII_distances}). The observed radial profile is measured up to $15.0_{-0.3}^{+0.4}$ kpc (Figure \ref{fig:source_radial_OII_FeII_MgII}). Furthermore, the observed radial profiles for Region U, Region E, and Region B reach radial distances of $\approx 5.7_{-0.2}^{+0.3}$  kpc, $8_{-0.2}^{+0.2}$ kpc, and $6.7_{-0.1}^{+0.1}$ kpc, respectively. (Figures \ref{fig:sourceOII_FeII_MgII_full_3EUB}, \& \ref{fig:src_radial_3EUB_OII_FeII_MgII}).
    
    \item We detect several \feii\ fine structure emission lines. Compared to the stellar continuum, the \feii\ fine structure emission is spatially extended (Figures \ref{fig:muse_OII_FeII_MgII_map_full}, \& \ref{fig:source_OII_FeII_MgII_full_CARC}). The measured maximum x- and y-extents after correcting for shear lensing+seeing in the source plane are $21.0_{-0.4}^{+0.4}$ kpc, and $13.7_{-0.3}^{+0.4}$ kpc, respectively (Figure \ref{fig:spatial_XY_extent_src} \& Table \ref{tab:CARC_MgII_distances}). The observed radial profile is extended up to a distance of $14.33_{-0.25}^{+0.27}$ kpc. In addition, the observed radial profiles for Region U, Region E, and Region B reach maximum radial extents of $\approx 9_{-0.2}^{+0.4}$ kpc, $9.2_{-0.2}^{+0.2}$ kpc, and $10_{-0.1}^{+0.1}$ kpc, respectively. (Figures \ref{fig:sourceOII_FeII_MgII_full_3EUB}, \& \ref{fig:src_radial_3EUB_OII_FeII_MgII}).
    
    \item The \mgii\ resonant emission lines $\lambda\lambda$ 2796, 2803 {\AA} are spatially extended in all regions in the image-plane and the source-plane. From the surface brightness radial profiles, we detect a patchy \mgii\ emission in the whole galaxy out to an average observed radial distance of $26.5_{-0.4}^{+0.5}$ kpc (with a maximum extent of $\Delta \delta = 30.0_{-0.5 }^{+0.7}$ kpc along the right ascension direction after correcting for the seeing and lensing shear) (Figures \ref{fig:source_OII_FeII_MgII_full_CARC}, \ref{fig:spatial_XY_extent_src}, \& \ref{fig:source_radial_OII_FeII_MgII}).
    
    \item After correcting for seeing and lensing shear, the \mgii\ emission profile is asymmetric and $20.9_{-0.6}^{+0.7}$\% more extended along the declination compared to the right ascension ($\Delta \delta$ or $\Delta y$ = $30.0_{-0.5 }^{+0.7}$ kpc, and $\Delta \alpha$ or $\Delta x$ = $24.8_{-0.5}^{+0.5}$ kpc; Figures \ref{fig:source_OII_FeII_MgII_full_CARC}, \ref{fig:spatial_XY_extent_src}, \& Table \ref{tab:CARC_MgII_distances}).
    
    \item We detect two distinct redshifted emission peaks at different velocities ($\Delta v \approx 400\ {\rm km\ s^{-1}}$) for the \mgii\ 2803 {\AA} emission line. The \mgii\ emission corresponding to these peaks is also spatially extended, similar to the full \mgii\ emission profile. The primary emission component is more extended along the declination direction by $\approx 2.5$ kpc after seeing+lensing shear correction. However, the secondary component shows more extent along the right ascension direction by $\approx 2.5$ kpc after correcting for the seeing+ lensing shear. (Figure \ref{fig:muse_mgII2803_primary_secondary_CARC} \& Table \ref{tab:CARC_MgII_distances}). This is an evidence for the complex, inhomogenous, and asymmetric nature of the geometry of galactic outflows.

    \item The kinematic offset ($\Delta v \approx 400\ {\rm km\ s^{-1}}$) and different morphology of the two \mgii\ 2803 {\AA} emission components suggest that these two emission peaks may be tracing two different parts of the outflowing gas. The outflowing gas may have different origin (e.g. different discrete star-bursts, or originating from different star-forming regions), although they appear approximately co-spatial in projection (Figures \ref{fig:MgII2803I_radial_x_y}, \& \ref{fig:MgII2803II_radial_x_y}).
    
    \item We detect \mgii\ emission in different of star-forming regions of the galaxy. Image 3 (tracing almost the whole galaxy) shows an observed radial extent of $24.5_{-1.6}^{+1.6}$ kpc. The observed radial extent of the \mgii\ emission is $9.6_{- 0.2 }^{+0.2 }$ kpc, $9.0_{-0.2}^{+0.3}$ kpc, and $7.4_{-0.1}^{+0.1}$ kpc at the $3\sigma$ significance for regions E, U, and B, respectively (Figures \ref{fig:sourceOII_FeII_MgII_full_3EUB} \& \ref{fig:src_radial_3EUB_OII_FeII_MgII}). This shows that individual star forming regions either started the outflows at different times, or they ejected the outflows at different velocities.
    \item The covering fraction $C_f(r)$ for {\feii} emission is higher than $C_f$ for {\oii} at all radial bins. In addition, $C_f$ approaches zero at 10 kpc for {\oii} and 15 kpc for {\feii}.
    
    \item We quantify the spatial covering fraction of \mgii\ emission, $C_f(r)$, the fraction of area on the sky (pixels) covered by significant \mgii\ emission ($>3\sigma$). We find that $C_f(r)$ is unity at $r < 3$ kpc, but rapidly falls off to $\sim$ 10\% at 20 kpc, with an excess of $C_f\sim 0.5$ at 9 kpc. This suggests a non-uniform and clumpy morphology of the outflowing gas. We characterize $C_f(r)$ with power law convolved with the seeing with index $\gamma = -1.25_{-0.02}^{+0.02}$. We quantify that the average \mgii\ emission $\langle C_f \rangle$ is $0.13_{-0.01}^{+0.01}$ deduced from the area of the outflow, within $r$=30 kpc (Figure \ref{fig:covering_fraction}).
\end{itemize}

For this galaxy, the fine-structure {\feii} emission shows spatial radial extent less than the spatial radial extent of the resonant {\mgii} emission in the counter arc. The nature of the spatial distribution of the fine-structure \feii\ emission in this galaxy suggests that \feii\ emission may not trace the optically thin parts of the galactic outflows. This is likely because the \feii\ is arising from fluorescence powered by resonant Fe II absorption \citep{prochaska2011simple}. Thus, the bulk of the \feii\ emission comes from the densest part of the outflow at the core region of the galaxy, and is not spatially scattered at large distances compared to {\mgii} emission. Using a galaxy sample from the MUSE Hubble Ultra Deep Field Survey, \cite{finley2017museSurvey} reported that low-mass galaxies with $< 10^9 M_{\odot}$ exhibit pure \mgii\ emission that may be tracing the star-forming H II regions (see also \citealt{chisholm2020optically}), whereas high-mass galaxies ($> 10^{10} M_{\odot}$) only exhibit \feii\ fluorescent emission without any \mgii\ emission, and intermediate mass galaxies exhibit both \feii\ and P-Cygni \mgii\ emission. \rcsa\ is a star-forming galaxy with $M_{\ast} = 6.3\pm 0.7\times 10^{9}M_{\odot}$ (log$_{10}(M_* / M_{\odot}) = 9.80 \pm 0.05$) \citep{wuyts2012stellar}. We detect both \feii\ and \mgii\ emission lines in our spectra. This is consistent with the intermediate mass regime ($ 9 < log_{10}(M_*/M_{\odot}) < 10$) case from \cite{finley2017museSurvey}. \cite{finley2017galactic} also detected fluorescent \feii\ emission (and no \mgii\ emission) in a star-forming galaxy at $z=1.29$ that is 70\% more extended than its stellar continuum out to $\sim$ 4 kpc. In our case, the {\feii}, and {\mgii} emission are spatially extended compared to the {\oii} nebular emission. 

Our main finding shows that \mgii\ emission halo around \rcsa\ has a size of $\sim$ 30 kpc. Although this emission is spatially extended for multiple different regions, the \mgii\ surface brightness profiles measured around individual star-forming regions of the galaxy are not uniform (Section \ref{sec:MgII_ems_src_plane}). These variations are significant, as they suggest that the outflowing gas traced by the \mgii\ emission is powered by the local star-forming regions in the host galaxy. Figure \ref{fig:source_OII_FeII_MgII_full_CARC} shows distinct structure in \mgii\ emission which strongly points towards a clumpy asymmetric outflow being driven from this galaxy. Together, these two pieces of evidence suggest that the properties of individual star-forming regions may determine how far outflowing gas can be driven from a galaxy \citep{bordoloi2016spatially}. We will explore this hypothesis in a follow-up paper that will study the kinematics of the outflowing gas in this galaxy in detail (Shaban et al. in prep).

Recently, new evidence has increasingly shown spatially extended emission owing to galactic outflows at different cosmic epochs. \cite{chen2021resolved} reported spatially extended $Ly\alpha$ emission around a group of three lensed galaxies at $z$ = 3.038 out to $\sim 30$ kpc. \cite{zabl2021muse} measured \mgii\ extended emission around a galaxy at $z=0.702$ up to $\approx 25$ kpc. One of the most extreme cases of spatially extended outflow was reported by \cite{rupke2019100}, who measured an outflow up to 100 kpc in a star-burst galaxy called `Makani' at redshift $\sim 0.47$ using \oii\ emission. This measurement is much larger than our nebular {\oii} emission extent. Recently, \cite{burchett2020circumgalactic} found that the \mgii\ (continuum-subtracted) emission around a star-forming galaxy TKRS4389 ($z$= 0.6942) extends out to a diameter of 31 kpc. Our findings of spatially extended \mgii\ emission around \rcsa\ to a maximum spatial extent of $\approx 30.0_{-0.5 }^{+0.7}$ kpc is comparable to these studies. Most of these studies (particularly with \mgii\ emission) have been at $z< 1$. Our finding is the highest redshift detection of spatially extended galactic wind traced by \mgii\ emission ($z \sim$ 1.7). Further, while most other studies have targeted unlensed galaxies to trace the global \mgii\ spatial extent, \rcsa\ is strongly gravitationally lensed, magnifying the \mgii\ emission in individual star-forming regions. This enables us to measure the spatial extent of the outflowing gas not only around the galaxy as a whole but also around individual star forming regions of the galaxy at $z\sim 1.7$.

Recent theoretical works have also reported how the CGM of galaxies will look in \mgii\ emission. \cite{nelson2021cold} reprocessed the TNG50 simulations to study the {\mgii} halos in the CGM of galaxies within redshift range $0.3 < z < 2$ and stellar mass range $7.5 < log (M_{*}/ M_{\odot}) < 11$. They assumed that {\mgii} emission is optically thin, and they neglect the impact of resonant scattering. They found out that the {\mgii} halos around the galaxies are ubiquitous in star-forming galaxies, regardless of the redshift or the stellar mass. One of the origins of these halos is galactic outflows with complex morphology. The measured covering fraction radial profile for {\rcsa} in this work is an indicator of the morphology and complexity of the outflow. It shows a variation in the morphology of the \mgii\ emission tracing the outflow. This is consistent with this simulation's results. Our 3$\sigma$ limits for Mg II emission surface brightness are $\mathrm{\sim 10^{-18}\ erg\ s^{-1} cm^{-2} arcsecond^{-2}}$. This is one order of magnitude higher than the surface brightness values of corresponding $z \sim 2$ galaxies in \cite{nelson2021cold}. If we scale our radial profile to match that of \cite{nelson2021cold} in the mass bins $M_{*} = 10^{9.5} M_{\odot}$ and $M_* = 10^{10} M_{\odot}$ at $z=2$, our observed {\mgii} emission radial profile is still more spatially extended than those reported in \cite{nelson2021cold}.

Our approach, that combines the power of IFUs and strong gravitational lensing is a powerful and promising way to study galactic outflows at high redshifts, given that a well-defined lens model exists. In the near future, by increasing the sample size, we aim to build up a robust and statistically significant sample of spatially resolved measurements of galactic outflows, that will significantly enhance our understanding of the morphology and spatial extent of the outflows at cosmic noon.

\acknowledgments
This work is based on observations collected at the European Organization for Astronomical Research in the Southern Hemisphere under ESO program 098.A-0459(A). In addition, we used observations made with the NASA/ESA Hubble Space Telescope, obtained from the data archive at the Space Telescope Science Institute (STScI). STScI is operated by the Association of Universities for Research in Astronomy, Inc. under NASA contract NAS 5-26555. A.S and R.B would like to thank Kavli Institute for Theoretical Physics (KITP), which is supported by the National Science Foundation (NSF) under Grant No. NSF PHY-1748958, for hosting the Fundamentals of Gaseous Halos workshop. S.L. acknowledges support by FONDECYT grant 1191232.

\vspace{5mm}
\facilities{MUSE-VLT}

\software{\texttt{Astropy} \citep{astropy2013astropy, astropy2018astropy}, \texttt{matplotlib} \citep{Hunter:2007} }

\appendix
\counterwithin{figure}{section}
\section{{\mgii} 2803 emission primary and secondary components radial profiles}\label{sec:MgII_components_extent}
\begin{figure*}
    \centering
    \includegraphics[width=\textwidth]{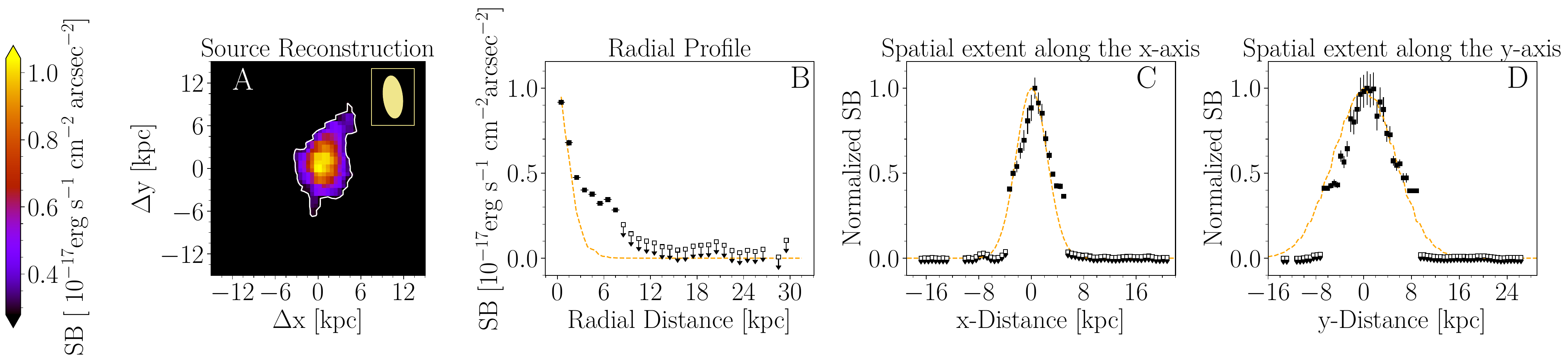}
    \caption{\textit{Panel A:} {\mgii} 2803 primary emission component surface brightness map in the source plane. The white contour correspond to 3$\sigma$ significance level. The yellow ellipse represents the shape and the size of the seeing in the source plane. \textit{Panel B:} Mean surface brightness radial profile. Panels \textit{C} and \textit{D} show the spatial extent of the surface brightness along the x-axis and y-axis, respectively. The filled black squares represent mean surface brightness greater than 3$\sigma$. The open black squares are 2$\sigma$ upper limits for bins with surface brightness below 3$\sigma$. The orange dashed lines in the panels represent seeing in the source plane.}
    \label{fig:MgII2803I_radial_x_y}
\end{figure*}

\begin{figure*}
    \centering
    \includegraphics[width=\textwidth]{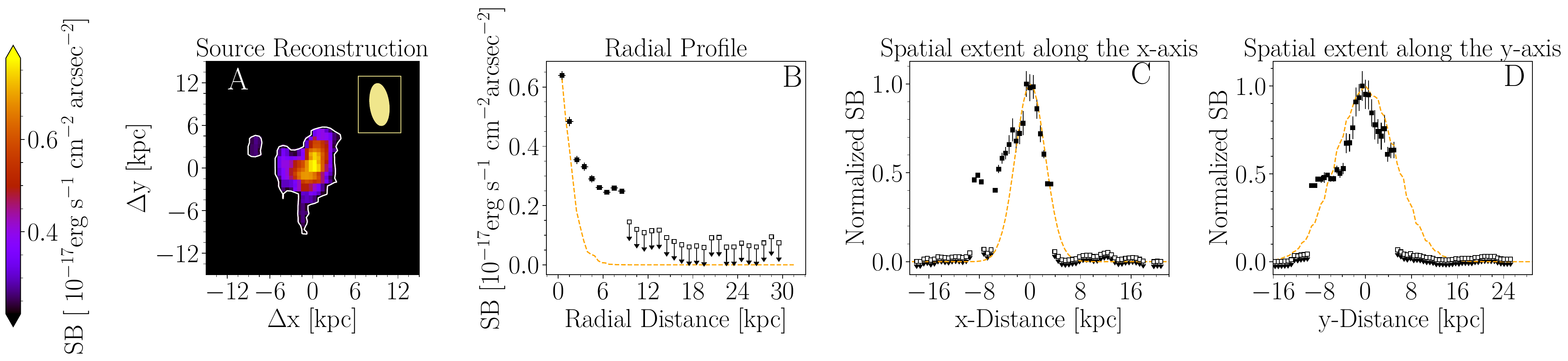}
    \caption{Same as Figure \ref{fig:MgII2803I_radial_x_y} but for the {\mgii} 2803 emission secondary component.}
    \label{fig:MgII2803II_radial_x_y}
\end{figure*}

In this appendix, we introduce the source plane surface brightness maps reconstructions,  the radial profiles, x- and y-spatial extents for the primary and secondary {\mgii} 2803 emission components in Figure \ref{fig:MgII2803I_radial_x_y} and Figure \ref{fig:MgII2803II_radial_x_y}, respectively.

\section{Radial Profiles and Covering fraction Parameters}\label{sec:appendix_radial_and_Cf}
\setcounter{table}{0}
\renewcommand{\thetable}{B.\arabic{table}}

\begin{figure*}
    \centering
    \includegraphics[width=\textwidth]{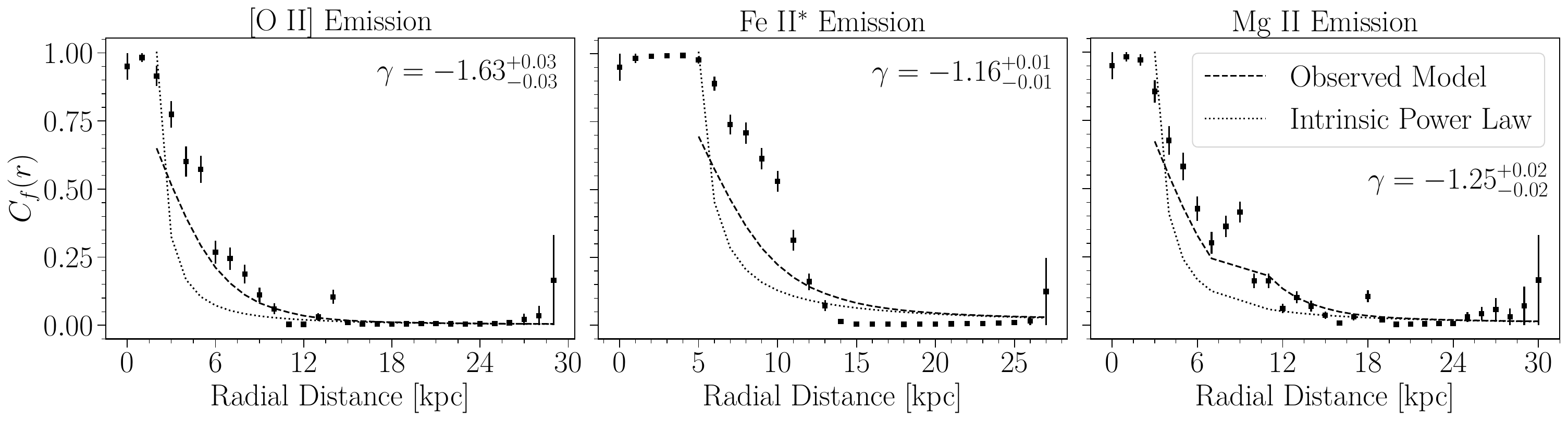}
    \caption{Modeling the covering fraction $C_f(r)$ as a function of radial distance for the {\oii} (\textit{Left}), {\feii} (\textit{Middle}), and {\mgii} emission (\textit{Right}) of the counter arc in the source plane. The dotted lines represent the intrinsic power law models used for the fitting. The dashed lines represent the observed model (intrinsic models after convolving them with the seeing) at the location of the counter arc in the source plane.}
    \label{fig:Cf_modeling_OII_FeII_MgII}
\end{figure*}

In this appendix, we report the best fit parameters of the \oii, \feii, and \mgii\ emission radial profiles models in Table \ref{table:radial_parameters}.1. The radial profile model is an exponential+power law in equation \ref{eq:radial_profile_model} in section \ref{sec:Method_Radial_Profiles}. We also show the fitting results for the emission covering fraction radial profiles for \oii, \feii, and {\mgii} in Figure \ref{fig:Cf_modeling_OII_FeII_MgII}.

\begin{deluxetable*}{cccccc}
\tablecolumns{6}
\tablewidth{0pt}
\tabletypesize{\footnotesize}
\tablecaption{The parameters for the \oii, \feii, and \mgii\ emission surface brightness} radial profiles models. The units for $I_{0,1}$ and $I_{0,2}$ are $\rm{erg \ s^{-1} \ cm^{-2} \ arcseconds^{-2}}$.
\label{table:radial_parameters}
\tablehead{
\colhead{Profile}&
\colhead{$I_{0,1}$ [$\times10^{-17}$]}&
\colhead{$I_{0,2}$ [$\times10^{-17}$]}&
\colhead{$r_0$ [kpc]}&
\colhead{$r_b$ [kpc]}&
\colhead{$\beta$}}
\startdata
Counter Arc   [O II] Emission & $0.41_{-0.01}^{+0.01}$ & $0.2_{-0.0}^{+0.0}$ & $2.05_{-0.04}^{+0.07}$ & $3.38_{-0.28}^{+0.52}$ & $-0.11_{-0.02}^{+0.01}$ \\ 
Image 3   [O II] Emission & $0.81_{-0.01}^{+0.01}$ & $0.21_{-0.0}^{+0.01}$ & $2.82_{-0.14}^{+0.12}$ & $3.65_{-0.47}^{+0.84}$ & $-0.16_{-0.02}^{+0.02}$ \\ 
Knot E   [O II] Emission & $0.64_{-0.03}^{+0.05}$ & $0.2_{-0.0}^{+0.0}$ & $1.16_{-0.08}^{+0.08}$ & $3.1_{-0.07}^{+0.14}$ & $-0.38_{-0.02}^{+0.02}$ \\ 
Knot U   [O II] Emission & $0.54_{-0.05}^{+0.04}$ & $0.11_{-0.01}^{+0.01}$ & $2.05_{-0.18}^{+0.17}$ & $3.57_{-0.43}^{+0.71}$ & $-0.41_{-0.08}^{+0.07}$ \\ 
Knot B   [O II] Emission & $0.44_{-0.03}^{+0.06}$ & $0.2_{-0.0}^{+0.0}$ & $0.57_{-0.05}^{+0.08}$ & $2.19_{-0.14}^{+0.22}$ & $-0.23_{-0.03}^{+0.03}$ \\ 
\hline
Counter Arc   Fe II$^*$ Emission & $2.95_{-0.18}^{+0.31}$ & $0.65_{-0.1}^{+0.14}$ & $2.06_{-0.04}^{+0.09}$ & $5.02_{-1.47}^{+1.8}$ & $-0.56_{-0.04}^{+0.05}$ \\ 
Image 3   Fe II$^*$ Emission & $1.99_{-0.08}^{+0.07}$ & $0.71_{-0.01}^{+0.01}$ & $4.32_{-0.17}^{+0.18}$ & $3.1_{-0.07}^{+0.15}$ & $-0.38_{-0.02}^{+0.01}$ \\ 
Knot E   Fe II$^*$ Emission & $2.39_{-0.43}^{+0.35}$ & $0.53_{-0.02}^{+0.05}$ & $1.9_{-0.12}^{+0.15}$ & $3.71_{-0.51}^{+0.77}$ & $-0.37_{-0.1}^{+0.1}$ \\ 
Knot U   Fe II$^*$ Emission & $1.89_{-0.24}^{+0.3}$ & $0.63_{-0.02}^{+0.04}$ & $2.4_{-0.28}^{+0.31}$ & $3.33_{-0.24}^{+0.42}$ & $-0.44_{-0.06}^{+0.07}$ \\ 
Knot B   Fe II$^*$ Emission & $1.82_{-0.24}^{+0.13}$ & $0.49_{-0.04}^{+0.06}$ & $1.24_{-0.18}^{+0.15}$ & $3.48_{-1.03}^{+1.0}$ & $-0.2_{-0.1}^{+0.09}$ \\ 
\hline
Counter Arc   Mg II Emission & $2.0_{-0.0}^{+0.0}$ & $0.4_{-0.0}^{+0.0}$ & $2.0_{-0.0}^{+0.0}$ & $4.09_{-0.07}^{+0.16}$ & $-0.07_{-0.0}^{+0.0}$ \\ 
Image 3   Mg II Emission & $2.92_{-0.04}^{+0.04}$ & $0.7_{-0.0}^{+0.0}$ & $2.04_{-0.03}^{+0.04}$ & $3.03_{-0.02}^{+0.04}$ & $-0.28_{-0.01}^{+0.01}$ \\ 
Knot E   Mg II Emission & $0.86_{-0.09}^{+0.08}$ & $0.41_{-0.0}^{+0.01}$ & $1.75_{-0.12}^{+0.14}$ & $3.38_{-0.28}^{+0.56}$ & $-0.16_{-0.04}^{+0.03}$ \\ 
Knot U   Mg II Emission & $1.17_{-0.12}^{+0.16}$ & $0.52_{-0.01}^{+0.03}$ & $2.33_{-0.27}^{+0.3}$ & $3.38_{-0.28}^{+0.57}$ & $-0.34_{-0.05}^{+0.06}$ \\ 
Knot B   Mg II Emission & $1.87_{-0.13}^{+0.09}$ & $0.45_{-0.01}^{+0.01}$ & $0.7_{-0.04}^{+0.06}$ & $3.42_{-0.97}^{+1.03}$ & $-0.01_{-0.02}^{+0.01}$ \\ 
\enddata
\vspace{-0.2cm}
\end{deluxetable*}

\bibliography{sample62.bbl} 

\end{document}